\documentclass[11pt]{article}
\pdfoutput=1
\usepackage{amsmath,amssymb,amsfonts}
\usepackage{cite}
\usepackage{graphicx}
\usepackage[colorlinks=true, pdfstartview=FitV, linkcolor=blue, citecolor=magenta, urlcolor=purple]{hyperref}
\usepackage{mdframed,framed}

\usepackage[usenames,dvipsnames]{xcolor}

\newcommand{\beq}{\begin{equation}}
\newcommand{\eeq}{\end{equation}}
\newcommand{\beqn}{\begin{eqnarray}}
\newcommand{\eeqn}{\end{eqnarray}}
\newcommand{\pa}{\partial}

\usepackage{cancel}

\newcommand{\comment}[1]{}

\newcommand{\D}{\text{d}}

\usepackage{tocloft}
\newlength{\leftfield}          
\newlength{\centerfield}        
\newlength{\rightfield}         
\newcommand{\twodigits}[1]{\ifnum#1<10 0\fi\number#1}
	
\newcommand{\mytitlebox}[4]{
	\framebox[\textwidth]{
		\begin{minipage}[c]{\textwidth}
		\makebox[\leftfield][l]{\littlefig{Physics-logo-small.png}{3} 
		}\ 
		\begin{minipage}[b]{\centerfield}\begin{center}{\bf\large #1}\\ #2\end{center}\end{minipage} 
		\begin{minipage}[b]{\rightfield}\begin{flushright}#3\\ Create: #4\\ Update: \the\year--\twodigits\month--\twodigits\day \end{flushright}\end{minipage} 
		\end{minipage}
	}
}	

\author{
  \begin{minipage}{.97\linewidth}
    \vspace{1cm}
       \begin{center}
      \begin{small}  
               \textbf{Luca Ciambelli},$^a$ 
             \textbf{Charles Marteau},$^b$ 
     \textbf{P. Marios Petropoulos}$^b$ and 
      \textbf{Romain Ruzziconi}$^a$
              \end{small}
    \end{center}
    \vspace{0.5cm}
      \hspace{2.4cm}\begin{minipage}{.7\linewidth}
\begin{center}     {\it \begin{footnotesize}
\hbox{
\kern-2cm
\vbox{
\begin{itemize}
 \item[$^a$]
Universit\'e Libre de Bruxelles \\and International Solvay Institutes\\
CP 231, 1050 Brussels, Belgium
      \end{itemize}
      \vskip0.20cm
      }   
\kern-2.8cm
      \vbox{
 \begin{itemize}
                  \item[$^b$]
Centre de Physique Th\'eorique -- CPHT\\ 
        Ecole Polytechnique, CNRS\footnote{\emph{Centre National de la Recherche Scientifique}, Unit\'e Mixte de Recherche UMR 7644.}\\
        Institut Polytechnique de Paris\\
        91128 Palaiseau Cedex, France   	   
      \end{itemize}}
    }
     \end{footnotesize}}
\end{center}
    \end{minipage}
      \vspace{0.5cm}
  \end{minipage}
}
\title{\vspace{1.5cm}
 \boldmath 
    \textbf{Fefferman--Graham and Bondi Gauges in the Fluid/Gravity Correspondence}
  \unboldmath
}

\date{}

\begin{document}

\begin{titlepage}
\maketitle
\thispagestyle{empty}

 \vspace{-12.cm}
  \begin{flushright}
  CPHT-PC031.052020
  \end{flushright}
 \vspace{11.7cm}

\begin{center}
\textsc{Abstract}\\  
\vspace{0.5cm}	
\begin{minipage}{1.0\linewidth}

In three-dimensional gravity, we discuss the relation between the Fefferman--Graham gauge, the Bondi gauge and the Eddington--Finkelstein type of gauge, often referred to as the derivative expansion, involved in the fluid/gravity correspondence. Starting with a negative cosmological constant, for each gauge, we derive the solution space and the residual gauge diffeomorphisms. We construct explicitly the diffeomorphisms that relate the various gauges, and establish the precise matching of their boundary data. We show that Bondi and Fefferman--Graham gauges are equivalent, while the fluid/gravity derivative expansion, originating from a partial gauge fixing,
exhibits an extra unspecified function that encodes the boundary fluid velocity. The Bondi gauge turns out to describe a subspace of the derivative expansion's solution space, featuring a fluid in a specific hydrodynamic frame. We 
pursue our analysis with the Ricci-flat limit of the Bondi gauge and of the fluid/gravity derivative expansion. The relations between them persist in this limit, which is well-defined and non-trivial. Moreover, the flat limit of the derivative expansion maps to the ultra-relativistic limit on the boundary. This procedure allows to unravel the holographic properties of the Bondi gauge for vanishing cosmological constant, in terms of its boundary Carrollian dual fluid.

\end{minipage}
\end{center}

\end{titlepage}

\begingroup
\hypersetup{linkcolor=black}
\tableofcontents
\endgroup
\noindent\rule{\textwidth}{0.6pt}
\newpage

\section{Introduction}

Three-dimensional vacuum Einstein gravity is a topological theory that provides an appropriate framework for addressing relevant questions about gravity. In this theory, physically different solutions of Einstein equations are labelled by their asymptotic charges, computed after specifying a set of boundary conditions. The latter are important because  their choice also dictates the associated asymptotic symmetries, and are set up in practice by specifying a particular gauge.\footnote{See e.g. \cite{Regge:1974zd,  Ashtekar:1981bq, Penrose:1982wp, Wald:1993nt, Wald:1999wa, Harlow:2019yfa, Herfray:2020rvq} and \cite{Troessaert:2013fma,Compere:2015knw, Donnay:2015abr, Hawking:2016msc, Averin:2016ybl, Grumiller:2016pqb, Detournay:2016sfv, Grumiller:2017sjh, oblak, Henneaux:2018cst, Gonzalez:2018jgp,  Parsa:2018kys,  Adjei:2019tuj, Barnich:2019vzx} 
for gauge-independent and gauge-dependent approaches, respectively.} Consequently, the study of different gauges -- and of their relationships -- becomes of primary importance. 

Depending on the problem one is addressing, different gauge fixings may be better suited. In the present work, we will focus on the Fefferman--Graham gauge, the Bondi gauge and the fluid/gravity derivative expansion. 
\begin{itemize}

\item \textit{Bondi gauge}. Introduced in \cite{Bondi1962, Sachs1962, Sachs:1962wk} for asymptotically flat spacetimes, it was revitalised in recent years due to a renewed interest in the symmetries of asymptotically flat spacetimes \cite{Barnich:2006av , Barnich:2010eb, Barnich:2016lyg} and in particular for their connections with soft theorems and memory effects \cite{He:2014laa, Campiglia:2014yka,  Strominger2017a, Ashtekar:2018lor, Compere:2018ylh, Donnay:2018neh, Adamo:2019ipt}. This gauge is implemented on a null direction, which makes it suited for the study of gravitational waves in higher dimensions. The Bondi gauge was also used to study asymptotically anti-de Sitter spacetimes and their flat limit in \cite{Barnich:2012aw, Barnich:2013sxa}, where a specific set of  boundary conditions was used. In \cite{Poole:2018koa , Compere:2019bua , Compere:2020lrt} (see also \cite{Ruzziconi:2019pzd}), less stringent conditions have been considered on the boundary geometric data allowing asymptotically locally anti-de Sitter and asymptotically locally flat spacetimes in four dimensions.\footnote{Here and in the following, ``locally'' means that the boundary-metric data are unspecified.}

\item \textit{Fefferman--Graham gauge}. This gauge was defined  in \cite{Feffe, Fefferman2011}, and has been  extensively used in holography \cite{Balasubramanian1999d, Skenderis2001}. It exhibits a radial direction (the holographic direction) parametrizing a family of time-like hypersurfaces, and radial evolution interpreted as the renormalization flow of the boundary theory. The bulk metric induces a conformal class of metrics and an energy--momentum tensor on the boundary.\footnote{See \cite{Ciambelli:2019bzz} for an explicit Weyl-covariant enhancement of this gauge.} The relationship between the Bondi and Fefferman--Graham gauges has been worked out for four-dimensional asymptotically locally AdS spacetimes in \cite{Poole:2018koa , Compere:2019bua}. 

\item \textit{Fluid/gravity derivative expansion}. This is an Eddington--Finkelstein type of gauge, which  appears in the fluid/gravity correspondence. The latter  set originally a relationship between boundary conformal relativistic fluids and bulk Einstein spacetimes \cite{Bhattacharyya2008a, Loganayagam2008, Haack:2008cp, Bhattacharyya2008b, Hubeny:2011hd}, and its name is borrowed from the derivative expansion used in the constitutive relations of fluid dynamics. The fluid/gravity derivative expansion (\emph{derivative expansion} for short) has been systematically studied beyond perturbation (i.e. in resummed forms) for anti de Sitter in
\cite{Caldarelli:2012cm,Mukhopadhyay:2013gja, Petropoulos:2014yaa,Gath:2015nxa, Petropoulos:2015fba}, and later generalized towards asymptotically flat spacetimes in three and four dimensions \cite{Ciambelli:2018wre, Campoleoni:2018ltl}. It is  implemented using a null bulk congruence, as is the Bondi gauge, though in a rather boundary-to-bulk spirit. It provides thus a concrete interpretation of the boundary data in terms of relativistic or more exotic Carrollian fluids, for asymptotically anti-de Sitter or flat spacetimes.

\end{itemize}

Our aim is here to deliver a comprehensive review of three-dimensional asymptotically locally AdS and asymptotically locally flat spacetimes. The presentation takes a slightly different and complementary perspective than the work \cite{CMPR}, following the pattern gauge-fixing, asymptotic behavior, solution space and its variation,  and residual diffeomorphisms. We build the coordinate transformations relating the above gauges, and set the precise dictionary between their solution spaces. A thorough description would require the computation of the associated surface charges, for which we set the stage.

In this analysis, we demand mild falloffs in order to allow for a conformal compactification. This minimal requirement is important in the spirit of finding the biggest solution space allowed for each gauge, including all possible arbitrary data with the most general corresponding asymptotic symmetries. Of course, our approach is meant to embrace restricted cases with stronger boundary conditions, as those studied in the quoted literature.

Performing the proposed study, and exhibiting the concrete diffeomorphisms that relate the various gauges is useful from several viewpoints. Firstly, these tools are necessary for comparing solution spaces, asymptotic symmetries, and ultimately surface charges.  Secondly, they reveal alternative interpretations to the associated data. 
In particular we will see through the comparison between the Bondi gauge and the derivative expansion, how to interpret the Bondi data in terms of the boundary relativistic conformal fluid. 
For example the bulk Bondi angular-momentum aspect turns out to be related to the boundary-fluid heat current, whereas the Bondi mass is encoded in its energy density. Lastly, this approach supplies the right frame for considering the flat limit -- whenever it is regular. This bulk limit induces an ultra-relativistic limit on the boundary,\footnote{Mathematical and physical queries on the ultra-relativistic limit, dubbed Carrollian, started with the work of L\'evy--Leblond \cite{Levy}, and gain attention over the recent years, often in conjunction with its dual Galilean counterpart 
\cite{Henneaux:1979vn, Duval:2014uoa, Duval:2014uva, Duval:2014lpa, Bekaert:2014bwa, Bekaert:2015xua, Hartong:2015xda, Figueroa-OFarrill:2018ilb, Morand:2018tke,Ciambelli:2019lap, Bergshoeff:2014jla, card, ba, CM1}.} 
which allows for a holographic description of the flat Bondi data. 
In the fluid/gravity language, this description relies on Carrollian hydrodynamics, along the lines developed in \cite{Ciambelli:2018xat}\footnote{Other methods for generalizing hydrodynamics on non-relativistic spacetimes can be found 
e.g. in \cite{Hassaine:1999hn, Horvathy:2009kz,deBoer:2017ing, Penna2017, Penna2018, Poovuttikul:2019ckt, deBoer:2020xlc}.} and successfully applied in \cite{Ciambelli:2018wre, Campoleoni:2018ltl} for making a step further in grasping asymptotically flat holography.\footnote{See for instance  
\cite{deBoer:2003vf,  Barnich:2012rz, Fareghbal:2013ifa, Strominger:2013jfa, Detournay:2014fva, Barnich:2015mui, Bagchi:2016bcd, Kapec:2016jld, Pasterski:2016qvg, Oblak:2016eij, Ball:2019atb, Merbis:2019wgk} for related work.}

In section \ref{AdS Gauges}, we define the gauges at hand in locally AdS$_3$ spacetime, derive their  general solution space, determine the residual gauge diffeomorphisms and work out the variations of the solution space. From this analysis, we observe that the solution spaces of Fefferman--Graham and Bondi gauges are parametrized by five functions, while the derivative expansion requires six functions. The additional function encodes the boundary-fluid velocity. The velocity of a relativistic fluid is admittedly redundant \cite{Landau1987Fluid, rezzolla2013relativistic} and its choice defines a \emph{hydrodynamic frame}. 
The explicit appearance of the latter in the solution space, already analyzed in \cite{Campoleoni:2018ltl} in relation with the conserved charges, shows that the corresponding redundancy is at best local. Notice that  even locally, the
 choice of hydrodynamic frame is bound to each specific physical situation. This feature has been discussed in the recent literature \cite{Kovtun:2012rj, Ciambelli:2017wou, Grozdanov:2019kge, Kovtun:2019hdm}.

In section \ref{Gauge Matching}, we design the diffeomorphisms relating these gauges. This allows to match explicitly the solution spaces and the residual gauge diffeomorphisms, extending thereby
 to three dimensions the results of \cite{Compere:2019bua} for Bondi and Fefferman--Graham. In particular, we show that the Bondi gauge appears as a sub-gauge of the derivative expansion, associated with a specific fluid frame. As anticipated, this endows the Bondi data with a fluid holographic interpretation. 

In section \ref{sec:Ricci Flat}, we focus on the Ricci-flat limit of the results established earlier for the Bondi gauge and the derivative expansion. Both admit a smooth limit, where the time-like boundary becomes null \cite{Barnich:2012aw, Barnich:2013sxa, Ciambelli:2018wre, Campoleoni:2018ltl}, as opposed to the Fefferman--Graham gauge. Furthermore, although the flat limit is straightforward in Bondi gauge, for the derivative expansion one needs to carefully define the behavior of the boundary fluid data at vanishing velocity of light. This is performed along the lines discussed in \cite{Ciambelli:2018wre, Campoleoni:2018ltl,Ciambelli:2018xat}, and guarantees 
that the bulk line element remains finite. As for $\text{AdS}_3$, the Bondi gauge is retrieved as a sub-gauge of a Ricci-flat fluid/gravity derivative expansion originally introduced in \cite{Ciambelli:2018wre, Campoleoni:2018ltl}. 

This work is accompanied with the Mathematica notebook  \texttt{Appendix.nb}, which gathers some lengthy expressions of the coordinate transformations discussed in section \ref{Gauge Matching}.

\section{Anti-de Sitter Gauges}
\label{AdS Gauges}

In this section we describe the advertised gauges, following a systematic pattern: we firstly define the gauge-fixing conditions for the metric, then we analyze the solution space of Einstein equations, and finally the variation of the latter under residual gauge diffeomorphisms.

\subsection{Fefferman--Graham}

The Fefferman--Graham description of asymptotically anti-de Sitter spacetimes is well-known \cite{Feffe, Fefferman2011}. We report it for completeness and for later comparison with the other gauges.

\subsubsection*{Definition}
In the Fefferman--Graham gauge, \cite{Feffe, Fefferman2011}, the metric is given by 
\begin{equation}
\D s^2 = \frac{\ell^2}{\rho^2} \D\rho^2 + g_{ab}(\rho, x) \D x^a \D x^b,
\label{gfFG}
\end{equation}
with coordinates $(\rho, x^a)$, $x^a = (t, \phi)$. The boundary is located at $\rho=0$ and $\ell$ is the AdS radius. The three gauge-fixing conditions are
\begin{equation}
g_{\rho \rho} = \frac{\ell^2}{\rho^2}, \quad g_{\rho a} = 0.
\label{gfFG fixing}
\end{equation}

Residual gauge diffeomorphisms $\xi$, namely diffeomorphisms that preserve the gauge-fixing conditions \eqref{gfFG fixing}, satisfy
\begin{equation}
\mathcal{L}_\xi g_{\rho \rho} = 0, \quad \mathcal{L}_\xi g_{\rho a} = 0.
\end{equation}
The explicit solution of these equations is given by 
\begin{equation}
\xi^\rho =  \rho \sigma (x), \quad \xi^a = \xi^a_0 (x) - \ell^2 \partial_b \sigma \int^\rho_0 \frac{\D \rho'}{\rho'} g^{ab} (\rho', x), 
\label{residual gauge diffeomorphisms FG}
\end{equation}
where $\sigma(x)$ and $\xi^a_0(x)$ are arbitrary functions of $x^a$.

\subsubsection*{Solution space}
\label{Solution space FG}

We impose the preliminary boundary condition $g_{ab} = \mathcal{O}(\rho^{-2})$. Solving the three-dimensional Einstein equations leads to the analytic finite expansion
\begin{equation}
g_{ab}(\rho , x)= \rho^{-2} g_{ab}^{(0)}(x) +  g_{ab}^{(2)}(x)  + \rho^2  g_{ab}^{(4)}(x),
\end{equation} where $g_{ab}^{(4)}$ is determined by $g_{ab}^{(0)}$ and $g_{ab}^{(2)}$ as
\begin{equation}
 g_{ab}^{(4)} = \frac{1}{4} g^{(2)}_{ac}g_{(0)}^{cd}g^{(2)}_{db}.
\end{equation} 
Einstein's equations leave $g^{(2)}_{ab}$ unspecified up to its trace $\text{Tr}\Big[g^{(2)}\Big]=-\frac{\ell^2}{2}R^{(0)}$ and the dynamical constraint $D_{(0)}^ag^{(2)}_{ab}=-\frac{\ell^2}{2}g^{(0)}_{ab}D^a_{(0)}R^{(0)}$. Here, $D^a_{(0)}$ is the covariant derivative with respect to $g^{(0)}_{ab}$ and indices are lowered and raised by $g^{(0)}_{ab}$ and its inverse. In the spirit of  \cite{Balasubramanian1999d, Skenderis2001}, we define the holographic energy--momentum tensor
\beqn
T_{ab} &=& \frac{1}{8 \pi G \ell} \left(g^{(2)}_{ab} - g_{ab}^{(0)} \text{Tr} \Big[g^{(2)}\Big]\right) \nonumber\\
&=& \frac{1}{8 \pi G \ell} \left(g^{(2)}_{ab} + \frac{\ell^2}{2} g_{ab}^{(0)}R^{(0)} \right).
\eeqn 
Therefore the Einstein equations infer
\begin{equation}
{T_a}^a =\frac{c}{24 \pi} R^{(0)} , \quad D^a_{(0)} T_{ab} = 0,
\label{trace condition}
\end{equation} where $c = \frac{3 \ell}{2 G}$ is the three-dimensional Brown--Henneaux central charge \cite{Brown:1986nw, Henningson1998}.  

The solution space is thus characterized by five arbitrary functions of $x^a$. Three are in the symmetric tensor $g_{ab}^{(0)}$ and two in the symmetric tensor $T_{ab}$ with constrained trace. These data are subject to two dynamical equations given by $D_{(0)}^a T_{ab} = 0$.

\subsubsection*{Variation of the solution space}

The residual gauge diffeomorphisms \eqref{residual gauge diffeomorphisms FG} evaluated on-shell are given by
\begin{equation}
\xi^\rho = \sigma \rho , \quad \xi^a = \xi^a_0 - \frac{\rho^2}{2} g_{(0)}^{ab} \ell^2 \partial_b \sigma + \frac{\rho^4}{4} g_{(0)}^{ac}g_{cd}^{(2)}g_{(0)}^{db}\ell^2 \partial_b \sigma + \mathcal{O}(\rho^6) .
\end{equation}
Under these residual gauge diffeomorphisms, the unconstrained part of the solution space transforms as\footnote{Our convention for the variation $\delta_\xi $ is the opposite of that used in \cite{CMPR}.} 
\begin{equation}
\delta_\xi g_{ab}^{(0)} = \mathcal{L}_{\xi_0} g_{ab}^{(0)} - 2 \sigma g_{ab}^{(0)},
\label{varmetric}
\end{equation} while the constrained part transforms as
\begin{equation}
\delta_\xi g_{ab}^{(2)} = \mathcal{L}_{\xi_0} g_{ab}^{(2)} - \frac{\ell^2}{2} \mathcal{L}_{\partial \sigma} g_{ab}^{(0)} ,
\end{equation}
from which one can extract the variation of $T_{ab}$.

A consistent boundary condition (that we will not impose in the subsequent developments), often used in the literature, is the Brown--Henneaux \cite{Brown:1986nw} condition: the boundary metric is frozen to be the flat metric $g_{ab}^{(0)}=\eta_{ab}$. When imposing this condition we recover the usual aymptotic symmetry group in AdS$_3$, i.e. the conformal group. Indeed, \eqref{varmetric} becomes
\begin{equation}
\delta_\xi \eta_{ab} = \mathcal{L}_{\xi_0} \eta_{ab} - 2 \sigma \eta_{ab}=0.
\end{equation}  
Tracing this last equation enables us to write $\sigma$ in terms of $\xi_0$. Therefore the symmetry algebra is uniquely specified by a vector $\xi_0$ that belongs to the boundary conformal algebra. Using this phase space, one can compute the associated surface charges and their algebra to deduce the Brown--Henneaux central charge.

\subsection{Bondi}

The Bondi gauge has been studied in e.g. \cite{Barnich:2006av, Barnich:2010eb}. We extend the analysis to asymptotically locally AdS spacetimes, including the boundary metric in the solution space. Similar results were obtained in four dimensions \cite{Compere:2019bua}. 

\subsubsection*{Definition}

In the Bondi gauge \cite{Bondi1962, Sachs:1962wk, Sachs1962, Barnich:2010eb, Barnich:2012aw}, the metric is given by
\begin{equation}
\D s^2 = \frac{V}{r} e^{2\beta} \D u^2 - 2 e^{2 \beta} \D u \D r + r^2 e^{2 \varphi} (\D \phi - U \D u)^2 ,
\label{Bondi metric}
\end{equation} with coordinates $(u, r, \phi)$. In this expression, $V$, $\beta$ and $U$ are functions of $(u, r, \phi)$, and $\varphi$ is function of $(u,\phi)$. The three gauge-fixing conditions are 
\begin{equation}
g_{rr} = 0, \quad g_{r \phi} = 0, \quad g_{\phi \phi} = r^2  e^{2 \varphi} .
\label{Bondi gauge fixing}
\end{equation} Note that $g_{\phi\phi} =  r^2 e^{2 \varphi}$ is the unique solution of the determinant condition
\begin{equation}
\partial_r \left(\frac{g_{\phi\phi}}{r^2} \right) =0,
\end{equation} which can be generalized to define the Bondi gauge in higher dimensions. 

The residual gauge diffeomorphisms $\xi$ preserving the Bondi gauge fixing \eqref{Bondi gauge fixing} have to satisfy the three conditions
\begin{equation}
\mathcal{L}_\xi g_{rr} = 0, \quad \mathcal{L}_\xi g_{r \phi} = 0, \quad g^{\phi\phi} \mathcal{L}_\xi g_{\phi \phi} = 2 \omega (u,\phi) .
\end{equation} The explicit solution of these equations is given by 
\beqn
\xi^u &=&  f, \label{resB1}\\
\xi^\phi &=& Y - \partial_\phi f\, e^{-2 \varphi} \int_r^{+\infty} \frac{\D r'}{{r'}^2} e^{2\beta} ,\label{resB2}\\
\xi^r &=& - r [ \partial_\phi \xi^\phi - \omega - U \partial_\phi f + \xi^\phi \partial_\phi \varphi + f \partial_u \varphi ],
\label{resB3}
\eeqn where $f(u,\phi)$, $Y(u,\phi)$ and $\omega (u,\phi)$ are arbitrary functions of $(u,\phi)$.

\subsubsection*{Solution space}
\label{Solution space}

This following analysis of  the general solution space in the gauge at hand 
generalizes the results of \cite{Barnich:2010eb}. There is no need of imposing any preliminary boundary condition here. This is in contrast with the procedure followed in the Fefferman--Graham gauge. Therefore, in three dimensions, the gauge conditions \eqref{Bondi gauge fixing} are to some extent stronger than those imposed for defining the Fefferman--Graham gauge \eqref{gfFG fixing}. 

First we impose the Einstein equations leading to the metric radial constraints. Solving $G_{rr} - \frac{1}{\ell^2} g_{rr} = R_{rr} = 0$ gives
\begin{equation}
\beta = \beta_0 (u, \phi).
\label{beta0}
\end{equation} 
Next, the equation $G_{r\phi} - \frac{1}{\ell^2} g_{r\phi} = R_{r\phi} = 0$ leads to
\begin{equation}
U= U_0(u, \phi) + \frac{1}{r}  2 e^{2\beta_0} e^{-2 \varphi} \partial_\phi \beta_0  - \frac{1}{r^2} e^{2\beta_0} e^{-2 \varphi} N(u, \phi).
\label{U}
\end{equation} 
Finally, $G_{u r} - \frac{1}{\ell^2} g_{ur} = 0$ gives
\begin{equation}
\frac{V}{r} = -\frac{r^2}{\ell^2} e^{2 \beta_0} - 2 r (\partial_u \varphi + D_\phi U_0 ) + M (u, \phi) + \frac{1}{r} 4 e^{2 \beta_0} e^{-2 \varphi} N \partial_\phi \beta_0 - \frac{1}{r^2} e^{2 \beta_0} e^{-2 \varphi} N^2,
\label{V}
\end{equation} where $D_\phi U_0 = \partial_\phi U_0 + \partial_\phi \varphi U_0$. Taking into account the previous results, the Einstein equation $G_{\phi\phi} - \frac{1}{\ell^2} g_{\phi \phi} =0$ is automatically satisfied at all orders. 

We now solve the Einstein equations in order to obtain the time evolution of $M$ and $N$. The equation $G_{u\phi} - \frac{1}{\ell^2}g_{u\phi} =0$ provides
\beqn
(\partial_u + \partial_u \varphi ) N &=& \left(\frac{1}{2} \partial_\phi + \partial_\phi \beta_0 \right) M -2 N \partial_\phi U_0 - U_0 (\partial_\phi N + N \partial_\phi \varphi )\nonumber\\
&& + 4 e^{2 \beta_0-2 \varphi} [2 (\partial_\phi \beta_0)^3 - (\partial_\phi \varphi) (\partial_\phi \beta_0)^2 + (\partial_\phi \beta_0 ) (\partial_\phi^2 \beta_0) ].
\eeqn
Moreover, $G_{uu} - \frac{1}{\ell^2}g_{uu} = 0$ imposes
\beqn
\partial_u M &=&(- 2 \partial_u \varphi + 2 \partial_u \beta_0-2\pa_\phi U_0+U_0 2\partial_\phi \beta_0- U_0 2\partial_\phi \varphi-U_0\partial_\phi)  M+ \frac{2}{\ell^2} e^{4 \beta_0-2 \varphi} [\partial_\phi N + N (4 \partial_\phi \beta_0 - \partial_\phi \varphi ) ] \nonumber\\
&&-2 e^{2 \beta_0-2 \varphi} \{\partial_\phi U_0 [8 (\partial_\phi \beta_0)^2 - 4 \partial_\phi \beta_0 \partial_\phi \varphi + (\partial_\phi \varphi)^2 + 4 \partial_\phi^2 \beta_0 - 2 \partial_\phi^2 \varphi ] -\partial_\phi^3 U_0 \nonumber\\
&&~~~~~~~~~~~~~~~~ + U_0 [ \partial_\phi \beta_0 (8 \partial_\phi^2 \beta_0 - 2 \partial_\phi^2 \varphi ) + \partial_\phi \varphi (- 2 \partial_\phi^2 \beta_0 + \partial_\phi^2 \varphi ) + 2 \partial_\phi^3 \beta_0 - \partial_\phi^3 \varphi ] \nonumber\\
&&~~~~~~~~~~~~~~~~ + 2 \partial_u \partial_\phi \beta_0 (4 \partial_\phi \beta_0 - \partial_\phi \varphi ) + \partial_u \partial_\phi \varphi (-2 \partial_\phi \beta_0 + \partial_\phi \varphi ) + 2 \partial_u \partial_\phi^2 \beta_0 - \partial_u \partial_\phi^2 \varphi \}.
\eeqn

The solution space is thus characterized by five arbitrary functions of $(u, \phi)$, given by $\beta_0$, $U_0$, $M$, $N$, $\varphi$, with two dynamical constraints expressing the time evolution of $M$ and $N$. This agrees with the results obtained when solving the Einstein equations in the Fefferman--Graham gauge. The precise matching between the two solution spaces will be established in section \ref{Fefferman-Graham and Bondi}. 

\subsubsection*{Variation of the solution space}

The residual gauge diffeomorphisms (\ref{resB1}--\ref{resB3}) evaluated on-shell are given by
\beqn
\xi^u &=& f,\label{resBON1}\\
\xi^\phi &=& Y - \frac{1}{r} \partial_\phi f\, e^{2\beta_0-2\varphi} ,\label{resBON2}\\
\xi^r &=& - r [ \partial_\phi Y - \omega - U_0 \partial_\phi f + Y \partial_\phi \varphi + f \partial_u \varphi ] \nonumber\\
&&+ e^{2\beta_0 - 2 \varphi} (\partial_\phi^2 f - \partial_\phi f \partial_\phi \varphi + 4 \partial_\phi f \partial_\phi \beta_0) - \frac{1}{r} e^{2\beta_0 - 2 \varphi}  \partial_\phi f\, N .\label{resBON3}
\eeqn
Under these residual gauge diffeomorphisms, the unconstrained part of the solution space transforms as
\beqn
\delta_\xi \varphi &=& \omega, \label{var1}\\
\delta_\xi \beta_0 &=& (f \partial_u + Y \partial_\phi)\beta_0 + \left(\frac{1}{2}\partial_u - \frac{1}{2} \partial_u \varphi + U_0 \partial_\phi \right) f - \frac{1}{2}(\partial_\phi Y + Y \partial_\phi \varphi - \omega ), \\
\delta_\xi U_0 &=& (f \partial_u + Y \partial_\phi - \partial_\phi Y ) U_0 - \left(\partial_u Y - \frac{1}{\ell^2} e^{4 \beta_0} e^{-2 \varphi} \partial_\phi f\right) + U_0 (\partial_u f +  U_0 \partial_\phi f) ,
\eeqn 
while the constrained part transforms as
\beqn
\delta_\xi N &=& (f \partial_u + Y \partial_\phi + 2 \partial_\phi Y +f \partial_u \varphi+  Y \partial_\phi \varphi -\omega-2 U_0 \partial_\phi f )N + M \partial_\phi f -e^{2\beta_0-2\varphi}[3\partial_\phi^2 f (2\partial_\phi \beta_0 - \partial_\phi \varphi)\nonumber \\
&& + \partial_\phi f( 4 (\partial_\phi \beta_0)^2 - 8 \partial_\phi \beta_0 \partial_\phi \varphi + 2 (\partial_\phi \varphi)^2 + 2 \partial_\phi^2 \beta_0 -\partial_\phi^2 \varphi)+ \partial_\phi^3 f  ], \\
\delta_\xi M &=&\frac{4}{\ell^2}\pa_\phi f e^{4\beta_0-2 \varphi}N+(\pa_u f+f \pa_u\varphi+\pa_\phi Y+Y\pa_\phi\varphi-\omega)M-2e^{2\beta_0-2\varphi}\Big[2\pa_\phi^2f\pa_u\beta_0+4\pa_u\pa_\phi f\pa_\phi\beta_0+\pa_u\pa_\phi^2 f \nonumber \\
&&+\pa_\phi^2f\pa_\phi U_0+8\pa_\phi^2 f\pa_\phi \beta_0 U_0+\pa_\phi f\Big((4\pa_\phi\beta_0-\pa_\phi\varphi)(2\pa_u \beta_0-\pa_u \varphi)+4\pa_u\pa_\phi\beta_0+\pa_\phi U_0(8\pa_\phi\beta_0-2\pa_\phi\varphi)\nonumber\\&&-\pa_\phi^2 U_0-2\pa_u\pa_\phi\varphi
+U_0(-4\pa_\phi\beta_0\pa_\phi\varphi+8(\pa_\phi\beta_0)^2+4\pa_\phi^2\beta_0+(\pa_\phi\varphi)^2-2\pa_\phi^2\varphi)\Big)-2\pa_\phi^2 f U_0\pa_\phi\varphi+\pa_\phi^3 fU_0\nonumber\\&&-\pa_u\pa_\phi f\pa_\phi\varphi-\pa_\phi^2 f\pa_u\varphi-2f\pa_\phi\beta_0\pa_u\pa_\phi\varphi+2\pa_\phi\beta_0\pa_\phi\omega-2\pa_\phi\beta_0\pa_\phi Y\pa_\phi\varphi-2\pa_\phi\beta_0\pa_\phi^2 Y-2\pa_\phi \beta_0 Y\pa_\phi^2\varphi\Big]\nonumber\\&&+f\pa_u M+\pa_\phi M Y. \label{var2}
\eeqn
These are the most general variations of the solution space in Bondi gauge. They are key ingredients in the computation of the asymptotic charge algebra. 

The conformal group is also obtained when imposing boundary conditions on the Bondi gauge in AdS. The Brown--Henneaux boundary condition $g^{(0)}_{ab}=\eta_{ab}$ is translated in Bondi gauge to
\begin{equation}
\varphi=0,\quad \beta_0=0, \quad U_0=0.
\end{equation}
The first one fixes $\omega$ to be zero, the second and third ones lead to two equations on $f$ and $Y$
\begin{equation}
\partial_u f=\partial_\phi Y, \quad \partial_u Y=\frac{1}{\ell^2}\partial_\phi f.
\end{equation}
The asymptotic symmetry algebra is therefore uniquely specified by a vector $\xi_0=f\partial_u+Y\partial_\phi$ that belongs to the conformal transformations of the flat boundary metric. 

\subsection{Derivative Expansion}\label{DEsec}

The fluid/gravity method for building Einstein spacetimes is not so different in spirit compared to Bondi's approach.  It is also an Eddington--Finkelstein type of gauge, not completely fixed though. Its solution space is determined by six functions, showing that this metric contains one additional arbitrary function. This means that the derivative expansion corresponds to a one-parameter family of true gauges, a particular member being the Bondi gauge. The solution space is mapped onto the data of a boundary relativistic fluid and the additional function is identified with the boundary fluid velocity. Since the boundary is two-dimensional, the orthogonal space to the fluid velocity is the one-dimensional space spanned by the Hodge dual of the velocity itself. The decomposition of tensors along these two directions, as already applied in \cite{Campoleoni:2018ltl, Banerjee:2014ita}, is a convenient property in two dimensions.

\subsubsection*{Definition and solution space}
 
The Eddington--Finkelstein type of gauge of the fluid/gravity holographic correspondence, the derivative expansion, is named after the standard derivative expansion used in fluid dynamics. The latter consists in expressing the various dissipative and non-dissipative quantities entering the relativistic-fluid energy--momentum tensor, as expansions in increasing derivative order of the fluid velocity (here $u^a$, normalized as $u^a \tilde g_{ab}u^b=-\tfrac{1}{\ell^2}$), the temperature and the chemical potentials (when extra currents are present). These are the fluid constitutive relations. 
In the original fluid/gravity correspondence, the bulk Einstein metric associated with a boundary relativistic fluid was set up order by order in inverse powers of the holographic coordinate $r$, which is a null radial coordinate \cite{Bhattacharyya2008a, Haack:2008cp, Bhattacharyya2008b, Hubeny:2011hd}. 
The coefficients of this expansion were derivatives of the fluid fundamental fields (velocity, temperature and chemical potentials) of increasing order,\footnote{More precisely, the starting point of the fluid/gravity advent was the exact Einstein spacetime generated by a boosted black brane. From there, new solutions were reached perturbatively,  corresponding to filling tubes centered around the elements of an Eddington--Finkelstein null radial congruence \cite{Bhattacharyya2008a}. The coefficients of this expansion were interpreted in terms of a boundary-fluid derivative expansion.} designed to ensure the invariance of the line element with respect to boundary Weyl transformations. 

As just summarized, this version of the fluid/gravity correspondence is restrictive because the class of Einstein spacetimes dual to fluid configurations is expected to be rather limited. In order to promote the fluid/gravity correspondence onto a genuine generating procedure for arbitrary Einstein spacetimes, we must set every quantity present in the energy--momentum tensor free and not determined by any sort of constitutive relation, accounting therefore for non-hydrodynamic modes. The fluid velocity and energy density, the heat current, the stress tensor, and the boundary metric become all  arbitrary functions, and these are the fundamental blocs that design the expansion in inverse powers of the radial light-like coordinate, dictated by Weyl covariance  \cite{Caldarelli:2012cm, Mukhopadhyay:2013gja, Petropoulos:2014yaa, Gath:2015nxa, Petropoulos:2015fba, Ciambelli:2018wre, Campoleoni:2018ltl, Ciambelli:2017wou}. Calling this expansion  a derivative expansion is a misnomer, which persist for convenience.

In three dimensions,  the most general line element expressed in the fluid/gravity derivative expansion, along the lines of generalization stated above spells as follows:
\begin{equation}
\D s^2= 2\ell^2 u_a \D x^a (\D r+rA_b \D x^b)+r^2 \tilde g_{ab}\D x^a \D x^b+8\pi G \ell^4 u_a \D x^a(\varepsilon u_b \D x^b+\chi \star u_b \D x^b)\label{DE},
\end{equation}
where $x^a=(x^0,\phi)=(\tfrac{u}{\ell},\phi)$. Since the boundary is two-dimensional, there exists a unique transverse direction to $u^a$, spanned by the boundary Hodge dual $\star u_a=\eta_{ab}u^b$.\footnote{Our conventions are: $\eta_{ab} = \sqrt{-\tilde g} \varepsilon_{ab}$ with $\varepsilon_{01} = +1$ and $\eta^{ab} = - \frac{1}{\sqrt{-\tilde g}} \varepsilon^{ab}$ with $\varepsilon^{01} = +1$. Hence, $\eta^{ab} \eta_{bc} = \delta^a_c$. Notice that the convention for the Hodge dual differs from \cite{Campoleoni:2018ltl, CMPR}.}  The various boundary tensors appearing in this line element are associated with the boundary fluid. In particular 
\beq
A_a=\ell^2(u^b\tilde{\nabla}_b u_a-\tilde{\nabla}_b u^b u_a)=\ell^2(\Theta^\star \star u_a-\Theta u_a),
\eeq
where we introduced the expansion of the fluid velocity $\Theta=\tilde\nabla_a u^a$ and of its dual $\Theta^\star=\tilde\nabla_a \star u^a$.  The vector $A_a$ can be interpreted as a Weyl connection \cite{Loganayagam2008}, whereas $\varepsilon$ and $\chi$ are the fluid energy density and heat current density, respectively. The latter are the local-equilibrium thermodynamic quantities entering the energy--momentum tensor of the boundary fluid:
\beq
\mathcal{T}_{ab}=2\ell^2\varepsilon u_a u_b+\varepsilon\tilde g_{ab}+\tau_{ab}+\ell^2 u_a q_b+\ell^2 u_b q_a.
\label{energy--momentum DE}
\eeq
In this expression $q_a=\chi\star u_a$ and $\tau_{ab}=\ell^2\tau \star u_a\star u_b$ are the dissipative tensors. 

The metric \eqref{DE}  was declared to be the most general in the following sense. According to the general pattern described earlier, for designing the fluid/gravity line element we use the fundamental blocks 
$\tilde g_{ab}$, $u_a$, $\star u_a$, $r$, $\varepsilon$ and $\chi$, which have weights $-2$, $-1$, $-1$, 1, 2 and 2. Weyl covariance is the guiding principle, but the final selection of the Weyl-covariant coefficients is operated by Einstein's equations involving the radial direction. This is how one is led to 
 \eqref{DE}, which is partly on-shell as metric \eqref{Bondi metric} is in Bondi gauge, once \eqref{beta0}, \eqref{U} and \eqref{V} are  taken into account. As already advertised, the fluid/gravity bulk reconstruction follows closely the scheme of the Bondi gauge -- particularly in three dimensions. It differs in the organisation principle of the bare terms appearing in the off-shell metric, here related to the dual fluid.

Now, we must impose that  \eqref{DE} obeys the transverse (with respect to $r$) Einstein equations, and this leads to the requirement that the boundary-fluid energy--momentum tensor satisfies
\beq
\tau=\frac{\tilde R}{8\pi G}, \qquad \tilde{\nabla}_a \mathcal{T}^{a}{}_{b}=-\tilde{\nabla}_a D^a{}_b,
\label{constraint on stress}
\eeq
where we introduced the traceless tensor
\beq
D_{ab}=\frac{\ell^4}{8\pi G}
\left[\left(u^c\tilde{\nabla}_c\Theta+ \star u^c \tilde{\nabla}_c\Theta^\star-\frac{1}{2\ell^2}\tilde R\right) 
\left(u_a u_b+
 \star u_a\star u_b
\right)-4\star u^c\tilde{\nabla}_c\Theta \ u_a \star u_b
\right].
\label{Dab def}
\eeq
The fluid characterized by $\mathcal{T}_{ab}$ evolves under the influence of a force $-\tilde\nabla_a D^{a}{}_{b}$. Projecting the equations of motion \eqref{constraint on stress} onto the velocity field and its dual, we obtain two scalar equations:
\beqn
u^a(\pa_a+2 A_a)\varepsilon+\star u^a(\pa_a+2 A_a)\chi &=&{\ell^2\over 4\pi G}\star u^a(\pa_a+2 A_a)(\star u^b\pa_b \Theta-u^b\pa_b \Theta^\star),\label{eom1}\\
\star u^a(\pa_a+2 A_a)\varepsilon+ u^a(\pa_a+2 A_a)\chi&=& 0.\label{eom2}
\eeqn

 The solution space is parametrized by six arbitrary functions of $x^a$. Three functions are encoded in $\tilde{g}_{ab}$, two are encoded in $\chi$ and $\epsilon$ with time evolution captured in \eqref{eom1} and \eqref{eom2}, and the last is in the normalized fluid velocity $u = u_a dx^a$ (in the following, we often take $u_\phi$ as the arbitrary function and $u_u$ is then completely determined through the normalization condition). Notice that there is one more arbitrary function than in the Fefferman--Graham and Bondi gauges. This is due to the fact that the derivative expansion provides only a partial gauge fixing \cite{Ruzziconi:2019pzd}. The extra degree of freedom in the parametrization of the solution space is the fluid velocity, as we will discuss in section \ref{Gauge Matching}. The hydrodynamic frame enters therefore explicitly (and by essence) the fluid/gravity approach, as opposed to the other gauges, which ignore the fluid. This latter fact does not imply that the fluid velocity is utterly unphysical. As established in \cite{Campoleoni:2018ltl}, one cannot dismiss the global outcome associated with the hydrodynamic frame, even though locally the choice of the fluid velocity is arbitrary.

\subsubsection*{On-shell residual gauge diffeomorphisms}

Since the derivative expansion is a partial gauge fixing, there is still an arbitrary function of $(r, x^a)$ in the residual gauge diffeomorphisms \cite{Ruzziconi:2019pzd}. In order to proceed, we start from the following ansatz:
\begin{equation}
\xi^a = \xi^a_{(0)} + \frac{1}{r} \xi^a_{(1)} , \quad \xi^r = r R + V + \frac{1}{r} W ,
\label{ansatz DE}
\end{equation} where $\xi^a_{(0)}= (F, \mathcal{Y})$, $\xi^a_{(1)} = (L, Z)$, $R$, $V$ and $W$ are functions of $x^a$. The form of these vectors is motivated by the expression of the residual gauge diffeomorphisms in the Bondi gauge (\ref{resBON1}-\ref{resBON3}) and the gauge matching that will be established in section \ref{Bondi and Derivative expansion}. The condition $\mathcal{L}_\xi g_{rr} = 0$ leads to
\begin{equation}
u_a \xi^{a}_{(1)} = 0 \quad \Longleftrightarrow \quad L = - \frac{u_\phi}{u_u} Z. 
\end{equation} Now, consider the condition $\mathcal{L}_\xi g_{ra} = \ell^2 H_a$ where $H_a$ are two functions of $x^a$ constrained by the fluid velocity normalization $\delta_\xi (\tilde{g}^{ab} u_a u_b) = 0$, or, equivalently,
\begin{equation}
\mathcal{L}_{\xi_{(0)}} \tilde{g}^{ab} u_a u_b + \frac{2}{\ell^2}  R +2 u^a H_a = 0 \quad \Longleftrightarrow \quad H_u = -\frac{1}{2 u^u} [ \mathcal{L}_{\xi_{(0)}} \tilde{g}^{ab}  u_a u_b + \frac{2}{\ell^2}  R + 2 u^\phi H_\phi]. 
\label{normalization Ha}
\end{equation} The requirement $\mathcal{L}_\xi g_{ra}|_{\text{order }1} = \ell^2 H_a$ gives
\begin{equation}
\xi^{(1)}_a = \ell^2 [ \mathcal{L}_{\xi^{(0)}} u_a + R u_a - H_a]
\label{xi1},
\end{equation} while $\mathcal{L}_\xi g_{ra}|_{\text{order }1/r^2} = 0$ yields
\begin{equation}
W= -4 \pi G \ell^2 \chi \xi^a_{(1)} \star u_a .
\end{equation} The remaining order $\mathcal{L}_\xi g_{ra}|_{\text{order }1/r} = 0$ does not impose further constraints on the parameters. Using $\delta_\xi A_a = \mathcal{L}_{\xi^{(0)}} A_a - u^c \tilde{\nabla}_c \xi^{(1)}_a + \tilde{\nabla}_c \xi_{(1)}^c u_a + \tilde{\nabla}_c u^c \xi^{(1)}_a - \xi_{(1)}^c \tilde{\nabla}_c u_a +\tilde{\nabla}_a R$, the condition $\mathcal{L}_\xi g_{ab}|_{\text{order }r} = 2 \ell^2 \delta_{\xi} (u_{(a} A_{b)})$ gives
\begin{equation}
V = - \tilde{\nabla}_a \xi^a_{(1)} .
\end{equation} Taking into account the previous results, the condition $\delta_\xi [\star u^a \star u^b (g_{ab}|_{\text{order }1}) ] = 0$ is straightforwardly satisfied. Similarly, using the equations of motion, $\mathcal{L}_\xi g_{ab}|_{\text{order }1/r} = 0$ can be shown to be automatically satisfied after a lengthy computation and using the equations of motion.

Hence, the family of residual gauge diffeomorphisms considered in \eqref{ansatz DE} is given explicitly by
\beqn
\xi^a &=& \xi^a_{(0)} + \frac{1}{r} \ell^2 [ \tilde{g}^{ab} \mathcal{L}_{\xi^{(0)}} u_b + R u^a - H^a] ,\label{resiDE1} \\
\xi^r &=& rR - \tilde{\nabla}_a \xi^a_{(1)} - \frac{4}{r} \pi G \ell^2 \chi \xi^a_{(1)} \star u_a \label{resiDE2} ,
\eeqn where the two functions $H_a = (H_u, H_\phi)$ are constrained through \eqref{normalization Ha}. 

These diffeomorphisms are parametrized by four arbitrary functions of $(u, \phi)$ given by $\xi^a_{(0)} = (F, \mathcal{Y})$, $R$ and $H_\phi$. Notice that there is an additional parameter $H_\phi$ compared to the residual gauge diffeomorphisms of the Bondi gauge (\ref{resBON1}-\ref{resBON3}). This generates the one-parameter family of diffeomorphisms enabling to change the fluid velocity. As we will see below, this is also the extra function needed to generate the diffeomorphism between the Bondi gauge and the derivative expansion (\ref{Bondi-DE for uphi non 0 1}-\ref{Bondi-DE for uphi non 0 2}).  Notice moreover that we can at any moment set consistently $u_\phi = 0$ and $H_\phi = 0$, and achieve the Bondi residual gauge diffeomorphisms. It is thus already obvious that the Bondi gauge defines a particular fluid frame from the derivative expansion viewpoint. We will elaborate on this later on.

\subsubsection*{Variation of the solution space}

Under the residual gauge diffeomorphisms (\ref{resiDE1}-\ref{resiDE2}) the unconstrained part of the solution space transforms as
\beqn
\delta_\xi \tilde{g}_{ab} &=& \mathcal{L}_{\xi^{(0)}} \tilde{g}_{ab} +2 R \tilde{g}_{ab} ,\\
\delta_\xi u_a &=& H_a,
\eeqn
while the constrained part transforms as\footnote{To obtain these variations, we have used various identities. In particular $
-u_a u_b +\star u_a\star u_b=\frac1{\ell^2}\tilde g_{ab}, \quad \Rightarrow \quad \star u^b\mathcal{L}_{\xi^{(1)}}u_b=0.$}
\beq
\delta_\xi \chi = \mathcal{L}_{\xi^{(0)}} \chi - 2 R \chi + 2 (\star u^a \xi_a^{(1)}) \varepsilon - \frac{1}{4 \pi G}[\tilde\nabla_a(\star u^a \tilde{\nabla}_c \xi^c_{(1)})+\ell^2u^c\tilde\nabla_c \xi^a_{(1)} u_a\Theta^\star+A_b\mathcal{L}_{\xi^{(1)}}\star u^b-\xi^c_{(1)}\tilde\nabla_c\Theta^\star ],
\label{var3.1}
\eeq
\beq
\delta_\xi \varepsilon =  \mathcal{L}_{\xi^{(0)}} \varepsilon - 2 R \varepsilon + 2 (\star u^a \xi_a^{(1)}) \chi + \frac{1}{4\pi G}[\tilde\nabla_a (u^a \tilde\nabla_b \xi^b_{(1)})+  \ell^2 u^c \tilde\nabla_c \xi^a_{(1)} u_a \Theta+A_a \mathcal{L}_{\xi^{(1)}} u^a-\xi^c_{(1)} \tilde\nabla_c \tilde\Theta ], 
\label{var3.2}
\eeq
where $\xi^{(1)}_a$ is expressed in \eqref{xi1} in terms of the gauge parameters. These two formulas are more compact than the variations of $N$ and $M$ obtained in the Bondi gauge \eqref{var2}. We will show that the latter can be obtained from \eqref{var3.1} and \eqref{var3.2} by setting $u_{\phi}=0$ and using the appropriate dictionary between the two gauges that we derive in the following section.

\section{Gauge Matching}
\label{Gauge Matching}

\subsection{From Bondi to Fefferman--Graham}
\label{Fefferman-Graham and Bondi}

\subsubsection*{Coordinate transformation}

Our aim is now to perform an explicit diffeomorphism relating the Bondi and Fefferman--Graham gauges. This is useful, among other things, for connecting their solution spaces. Following \cite{Poole:2018koa, Compere:2019bua}, we proceed in two steps.
We move firstly from Bondi to tortoise coordinates $(u,r,\phi) \to (t^*,r^*,\phi^*)$,
\begin{equation}
u = t^* - r^*, \quad r = - \ell \cot \left(\frac{r^*}{\ell} \right), \quad \phi^* = \phi,
\end{equation}
and secondly to Fefferman--Graham $(t^*,r^*,\phi) \to (\rho,t,\phi)$, with
\beqn
t^* &=& t + T_1(t,\phi) \rho +  T_2(t,\phi) \rho^2 + T_3 (t, \phi) \rho^3 + \mathcal{O}(\rho^4), \label{diffeo Bondi FG 1} \\
r^* &=& R_1(t,\phi) \rho + R_2(t,\phi)\rho^2 + R_3(t, \phi) \rho^3 + \mathcal{O}(\rho^4),\\
\phi^* &=& \phi + Z_1(t,\phi) \rho + Z_2 (t,\phi) \rho^2 + Z_3 (t, \phi) \rho^3 + \mathcal{O}(\rho^4).
\label{diffeo Bondi FG 2}
\eeqn 
The functions $T_i (t,\phi)$, $R_i(t,\phi)$ and $Z_i (t,\phi)$ ($i=1,2,3$) can be worked out explicitly.\footnote{We report them in the attached Mathematica file \texttt{Appendix.nb}.} For the sake of brevity, we report here only the leading orders
\beqn
R_1(t, \phi) &=& \ell^2,\\
R_2 (t, \phi) &=& - e^{-2\beta_0}\ell^4(\partial_\phi U_0 + U_0 \partial_\phi \varphi + \partial_t \varphi ), \\
&& \nonumber \\
T_1(t,\phi) &=& \ell^2 (1 - e^{-2\beta_0}), \\
T_2(t,\phi) &=& -e^{-4\beta_0} \ell^4 [e^{2\beta_0}\partial_\phi U_0 + U_0 (\partial_\phi\beta_0 + e^{2\beta_0} \partial_\phi \varphi) + \partial_t \beta_0 + e^{2\beta_0} \partial_t \varphi ], \\
&& \nonumber \\
Z_1 (t,\phi) &=&  - \ell^2 e^{-2\beta_0} U_0, \\
Z_2 (t,\phi) &=& \frac{1}{2}\ell^2 [2 e^{-2\varphi} \partial_\phi \beta_0 - 2 e^{-4\beta_0} \ell^2 U_0^2 \partial_\phi \beta_0 + e^{-4\beta_0}\ell^2 \partial_t U_0+ e^{-4\beta_0} \ell^2 U_0 (\partial_\phi U_0 - 2 \partial_t \beta_0)].
\eeqn

\subsubsection*{Solution space matching}

Using the diffeomorphism (\ref{diffeo Bondi FG 1}-\ref{diffeo Bondi FG 2}), the solution space of the Fefferman--Graham gauge (left-hand side) is related to the solution space of the Bondi gauge (right-hand side) through 
\begin{equation}
g^{(0)}_{ab} = \begin{pmatrix}
-\frac{e^{4\beta_0}}{\ell^2}+ e^{2\varphi} U_0^2 &-e^{2\varphi} U_0 \\
-e^{2\varphi} U_0 &e^{2\varphi} 
\end{pmatrix} 
\end{equation} and 
\beqn
T_{tt} &=& \frac{1}{16\pi G \ell} e^{-4\beta_0 - 2 \varphi} \{4e^{8 \beta_0} [2 (\partial_\phi \beta_0)^2 - \partial_\phi \beta_0 \partial_\phi \varphi + \partial_\phi^2 \beta_0 ] + e^{4 \beta_0 + 2 \varphi}[ e^{2 \beta_0} (M-4 N U_0 )\nonumber\\
&&- \ell^2 ( (\partial_\phi U_0)^2 + U_0^2 (-8 \partial_\phi \beta_0 \partial_\phi \varphi + (\partial_\phi \varphi)^2 + 4 \partial_\phi^2 \varphi)+ (\partial_t \varphi)^2 + 2 \partial_\phi U_0 (U_0 (-4 \partial_\phi \beta_0 + 3 \partial_\phi \varphi )+ \partial_t \varphi) \nonumber\\
&&+ 2 U_0 (2 \partial_\phi^2 U_0 + (-4 \partial_\phi\beta_0 + \partial_\phi \varphi) \partial_t \varphi + 2 \partial_t \partial_\phi \varphi ))]+ e^{4 \varphi} \ell^2 U_0^2 [ e^{2\beta_0} M + \ell^2 ( (\partial_\phi U_0)^2 \nonumber\\
&&+U_0^2 (- 4 \partial_\phi \beta_0 \partial_\phi \varphi + (\partial_\phi \varphi)^2 + 2 \partial_\phi^2 \varphi )    +2 \partial_\phi \varphi \partial_t U_0 + \partial_t \varphi (-4 \partial_t \beta_0  + \partial_t \varphi) + 2 \partial_\phi U_0 (2 U_0 (- \partial_\phi \beta_0 + \partial_\phi \varphi ) \nonumber \\
&&- 2 \partial_t \beta_0+ \partial_t \varphi ) + 2 U_0 (\partial_\phi^2 U_0 - 2 \partial_\phi \beta_0 \partial_t \varphi + \partial_\phi \varphi ( -2 \partial_t \beta_0 + \partial_t \varphi ) + 2 \partial_t \partial_\phi \varphi) + 2 (\partial_t \partial_\phi U_0 + \partial_t^2 \varphi ) )] \} ,
\eeqn
\beqn
T_{t \phi} &=& \frac{1}{16\pi G \ell}e^{-4\beta_0}\{ 2e^{6 \beta_0} N - 2 e^{4\beta_0} \ell^2 [\partial_\phi U_0 (2 \partial_\phi \beta_0 - \partial_\phi \varphi) - \partial_\phi^2 U_0 + U_0 (2 \partial_\phi \beta_0 \partial_\phi \varphi - \partial_\phi^2 \varphi ) + 2 \partial_\phi \beta_0  \partial_t \varphi\nonumber \\
&&- \partial_t \partial_\phi \varphi ] + e^{2 \varphi} \ell^2 U_0 [ - e^{2 \beta_0} M - \ell^2 ( (\partial_\phi U_0)^2 + U_0^2 (-4\partial_\phi \beta_0 \partial_\phi \varphi + (\partial_\phi \varphi )^2 + 2 \partial^2_\phi \varphi) \nonumber\\
&&+ 2 \partial_\phi \varphi \partial_t U_0 + \partial_t \varphi (-4 \partial_t \beta_0 + \partial_t \varphi ) + 2 \partial_\phi U_0 ( 2 U_0 (-\partial_\phi \beta_0 + \partial_\phi \varphi ) -2 \partial_t \beta_0 + \partial_t \varphi) \nonumber\\
&&+ 2 U_0 (\partial_\phi^2 U_0 - 2 \partial_\phi \beta_0 \partial_t \varphi + \partial_\phi \varphi ( - 2 \partial_t \beta_0 + \partial_t \varphi) +2 \partial_t \partial_\phi \varphi) + 2 (\partial_t \partial_\phi U_0 + \partial_t^2 \varphi ))] \} ,
\eeqn
\beqn
T_{\phi \phi} &=& \frac{1}{16\pi G\ell} e^{-4 \beta_0+ 2 \varphi} \{ e^{2 \beta_0} \ell^2 M + \ell^4 [ (\partial_\phi U_0)^2 + U_0^2 (-4 \partial_\phi \beta_0 \partial_\phi \varphi + (\partial_\phi \varphi)^2 + 2 \partial_\phi^2 \varphi )+ 2 \partial_\phi \varphi \partial_t U_0\nonumber \\
&&+ \partial_t \varphi (-4 \partial_t \beta_0 + \partial_t \varphi ) + 2 \partial_\phi U_0 (2 U_0 (-\partial_\phi \beta_0 + \partial_\phi \varphi) - 2 \partial_t \beta_0 + \partial_t \varphi ) + 2 U_0 (\partial_\phi^2 U_0 - 2 \partial_\phi \beta_0 \partial_t \varphi \nonumber\\
&&+\partial_\phi \varphi (-2 \partial_t \beta_0 + \partial_t \varphi) + 2 \partial_t \partial_\phi \varphi ) + 2 (\partial_t \partial_\phi U_0 + \partial_t^2 \varphi )] \}.
\eeqn
One can check on the right-hand-side expressions that the trace condition given by the first equation of \eqref{trace condition} is satisfied.

Taking $U_0 = 0$, $\beta_0 = 0$ and $\varphi = \bar{\varphi}$ (three-dimensional analogue of the boundary gauge fixing of \cite{Compere:2019bua}), $T_{ab}$ reduces to 
\begin{equation}
T_{ab} = \frac{1}{16\pi G \ell} \begin{pmatrix}
M - \ell^2 (\partial_t \bar{\varphi})^2 &2N + 2 \ell^2 \partial_t \partial_\phi \bar{\varphi}\\
2N + 2 \ell^2 \partial_t \partial_\phi \bar{\varphi} &e^{2\bar{\varphi}} \ell^2 [M + \ell^2 ((\partial_t \bar{\varphi})^2 + 2 \partial_t^2 \bar{\varphi} )]  
\end{pmatrix} .
\end{equation}
It would be interesting to pursue the study of various boundary gauge-fixed energy--momentum tensors in terms of Bondi data.

\subsubsection*{Residual gauge parameters matching}

Using the diffeomorphism (\ref{diffeo Bondi FG 1}-\ref{diffeo Bondi FG 2}), the parameters of the residual gauge diffeomorphisms of the Fefferman--Graham gauge \eqref{residual gauge diffeomorphisms FG} are related to those of the Bondi gauge (\ref{resB1}--\ref{resB3}) as\footnote{The matching is obtained comparing the leading order of the transformed residual gauge diffeomorphisms only. To map the full residual vector of one gauge to the full residual vector of the other, one needs to take into account the metric-fields dependence of the diffeomorphism -- see \cite{Compere:2016hzt}.}
\beqn
\xi_0^t &=& f, \\
\xi_0^\phi &=& Y, \\
\sigma &=& \partial_\phi Y - \omega  - U_0 \partial_\phi f + Y \partial_\phi \varphi + f \partial_t \varphi .
\eeqn

\subsection{From Bondi to Derivative Expansion}
\label{Bondi and Derivative expansion}

We would like now to set up an important result, which is the bridge between the Bondi gauge and  the derivative expansion. We firstly show that the Bondi solution space is actually embedded in that of the derivative expansion, and then explicitly construct the diffeomorphism that enables us to pass from the one to the other.

\subsubsection*{Bondi gauge as a sub-gauge of the derivative expansion}

For $u_\phi =0$, which is the additional constraint needed for a definite gauge fixing in the derivative expansion, the following identifications apply between the derivative-expansion and Bondi solution spaces:
\beqn
u_u &=& - \frac{1}{\ell^2} e^{2\beta_0}, \label{IndentBDE 1} \\
\tilde{g}_{\phi\phi} &=& e^{2\varphi}, \\
\tilde{g}_{uu} &=&- \frac{e^{4\beta_0}}{\ell^2} + e^{2\varphi} U_0^2, \\
\tilde{g}_{u\phi} &=& - e^{2\varphi} U_0, \\
\chi &=& \frac{1}{4\pi G \ell} e^{-\varphi} N ,
\label{chiN} \\
\varepsilon &=& \frac{1}{8\pi G} (e^{-2 \beta_0}M + 4 e^{-2 \varphi} (\partial_\phi \beta_0)^2 )  . \label{IndentBDE 2} 
\eeqn
The Bondi gauge turns out to be equivalent to the derivative expansion, locked in a specific hydrodynamic frame corresponding to $u_\phi =0$. Different frames are encoded in different values of $u_\phi$, which is the extra parameter present in the derivative expansion. The last two equations, \eqref{chiN} and \eqref{IndentBDE 2} show that the heat density $\chi$ is proportional to the angular momentum aspect $N$, while the fluid energy density $\varepsilon$ is simply related to the Bondi mass $M$.\footnote{This property is generically valid in higher dimension. In three dimensions, this interpretation of the boundary-fluid data was discussed in \cite{Campoleoni:2018ltl}, in relation with the bulk surface charges.}   

\subsubsection*{Coordinate transformation}

We now match the Bondi gauge to the derivative expansion letting the fluid velocity free (i.e. $u_\phi \neq 0$). Let us perform the following diffeomorphism from Bondi to derivative expansion:
\beqn
u_{B} &=& u + T_1[u,\phi;u_\phi(u,\phi)] r^{-1} +  T_2[u,\phi;u_\phi(u,\phi)]  r^{-2} + T_3 [u,\phi;u_\phi(u,\phi)]  r^{-3} + \mathcal{O}( r^{-4}), \label{Bondi-DE for uphi non 0 1}\\
r_{B}^{-1} &=& r^{-1} + R_2[u,\phi;u_\phi(u,\phi)] r^{-2} + R_3[u,\phi;u_\phi(u,\phi)]  r^{-3} + \mathcal{O}(r^{-4}),\\
\phi_{B} &=& \phi + Z_1[u,\phi;u_\phi(u,\phi)]  r^{-1} + Z_2 [u,\phi;u_\phi(u,\phi)]   r^{-2} + Z_3 [u,\phi;u_\phi(u,\phi)]  r^{-3} + \mathcal{O}( r^{-4}),
\label{Bondi-DE for uphi non 0 2}
\eeqn
where $(u_B, r_B, \phi_B)$ are the Bondi coordinates and $(u,r,\phi)$ are the derivative expansion coordinates. Notice that the functions of the diffeomorphisms depend on one parameter $u_\phi(u,\phi)$, and are all identically zero when $u_\phi = 0$ since we are already in the Bondi gauge in this case. For the sake of completeness, we report here the leading order of the diffeomorphisms,\footnote{The subleading functions are in the Mathematica file \texttt{Appendix.nb}.} defining $\mathcal{N}=\sqrt{1+e^{-2\varphi}\ell^2 u_\phi^2}$ (for $u_\phi=0$, $\mathcal{N}=1$)
\beqn
T_1(u,\phi)&=&-e^{-2\beta_0} \ell^2 \left(\mathcal{N}-1\right),\\
Z_1(u,\phi)&=&-e^{-2(\beta_0+\varphi)}\ell^2 \left(e^{2 \beta_0}u_\phi+e^{2 \varphi}U_0(\mathcal{N}-1)\right)\\
R_2(u,\phi)&=&-\frac{\ell^2}{\mathcal{N}} e^{-2 (\varphi+\beta_0)} \Big(2 e^{2 \beta_0} \mathcal{N} u_\phi \pa_\phi\beta_0+e^{2 \beta_0} \mathcal{N} \pa_\phi u_\phi-e^{2 \beta_0} \mathcal{N} u_\phi \pa_\phi\varphi +U_0 \left(\ell^2 u_\phi\pa_\phi u_\phi-(\mathcal{N}-1) e^{2 \varphi } \pa_\phi\varphi\right)\nonumber \\
&&+\pa_\phi U_0 \left(\ell^2 u_\phi^2-(\mathcal{N}-1) e^{2 \varphi
   }\right)+\ell^2 u_\phi\pa_u u_\phi - e^{2 \varphi } \pa_u\varphi(\mathcal{N}-1)\Big).
\eeqn

\subsubsection*{Solution space matching}

We are now ready to make the following identification between the solution space of the derivative expansion (left-hand side) and of the Bondi gauge (right-hand side):
\beqn
u_u &=& -U_0 u_\phi - \frac{1}{\ell^2} e^{2\beta_0 - \varphi} \sqrt{e^{2\varphi} + \ell^2 u_\phi^2} , \label{MathchingBDE 1}\\
\tilde{g}_{\phi\phi} &=& e^{2\varphi}, \\
\tilde{g}_{uu} &=& -\frac{1}{\ell^2} e^{4\beta_0} + e^{2\varphi} U_0^2 ,\\
\tilde{g}_{u\phi} &=& - e^{2 \varphi} U_0 . \label{MathchingBDE 2}
\eeqn
The expressions of $\chi$ and $\epsilon$ in terms of the Bondi solution space are lengthy and we do not write them explicitly -- they can be found in the attached Mathematica file \texttt{Appendix.nb} written explicitly in the sub-case $U_0=0=\beta_0$. The expressions of $u_u$, $\tilde{g}_{ab}$, $\chi$ and $\epsilon$ in terms of the Bondi functions depend on the parameter $u_\phi(u,\phi)$. They reduce to (\ref{IndentBDE 1}-\ref{IndentBDE 2}) when $u_\phi = 0$, as they should. As anticipated, the infinitesimal form of (\ref{Bondi-DE for uphi non 0 1}-\ref{Bondi-DE for uphi non 0 2}) is generated by $H_\phi$, the function introduced in section \ref{DEsec} as generator of residual gauge diffeomorphism. This is consistent, since it is indeed $H_\phi$ that generates $u_\phi$.

\subsubsection*{Residual gauge parameters matching}

The parameters of the residual gauge diffeomorphisms of the derivative expansion are related to those of the Bondi gauge as
\beqn
F &=& f, \\
\mathcal{Y} &=& Y, \\
R &=&  - \partial_\phi Y + \omega + U_0 \partial_\phi f - Y \partial_\phi \varphi - f \partial_u \varphi ,
\eeqn and $H_\phi$ has no equivalent in Bondi gauge, since it is associated with a shift of $u_\phi$, zero in the latter.

\subsection{Derivative Expansion and Fefferman--Graham}

\subsubsection*{Solution space matching}

The solution space of Bondi gauge has been identified with the solution space of Fefferman--Graham gauge and the solution space of derivative expansion. Therefore, this automatically leads to an identification of the solution spaces between Fefferman--Graham (left-hand side) and derivative expansion (right-hand side), without further computation:
\beqn
g^{(0)}_{ab} &=& \tilde{g}_{ab} ,\\
T_{ab} &=&{\ell\over 2} (\mathcal{T}_{ab} + D_{ab} ),
\label{EMmatching}
\eeqn where $\mathcal{T}_{ab}$ is the energy--momentum tensor \eqref{energy--momentum DE} and $D_{ab}$ is given in \eqref{Dab def}.

Recalling that the derivative expansion has six independent parameters while the Fefferman--Graham gauge has only five, we see that the various degrees of freedom have been shuffled in moving from one gauge to the other. The first equation matches the same number of pieces of data on both side, while the second matches two pieces of data on the left (the tensor $T_{ab}$ is symmetric and traceless) with three pieces of data on the right ($\varepsilon$, $\chi$ and the normalized velocity $u^a$). We conclude that we have a redundant description of the Fefferman--Graham data in terms of the fluid ones. To fix this redundancy one has to make a choice of fluid frame, which is exactly what we did to recover the Bondi gauge. Another way to verify the validity of equation \eqref{EMmatching} is to compute the holographic energy--momentum tensor of the derivative expansion using the Balasubramanian--Kraus method, that we know gives exactly $T_{ab}$ in the Fefferman--Graham gauge. This was done in \cite{Campoleoni:2018ltl}.

\subsubsection*{Residual gauge parameters matching}

The parameters of the residual gauge diffeomorphisms of the Fefferman--Graham gauge are related to those of the derivative expansion as
\beqn
\xi^t_0 &=& F, \\
\xi^\phi_0 &=& \mathcal{Y}, \\
\sigma &=&  - R ,
\eeqn and $H_\phi$ has no equivalent, as $u_\phi$ does not appear in Fefferman--Graham gauge.

\subsection{Holographic-Fluid Interpretation}
\label{Holographic Fluid Interpretation}

As we have seen, for asymptotically locally three-dimensional AdS spacetimes, the Bondi gauge is reached from the fluid/gravity side by setting $u_\phi=0$ in the derivative expansion. The relativistic velocity vector is then given by
\beq
 u^a\partial_a = e^{-2\beta_0}\left(\pa_u+U_0\pa_\phi\right),
\eeq
whereas, as observed in  \eqref{chiN} and \eqref{IndentBDE 2}, the heat density is proportional to the angular momentum aspect and the fluid energy density is related to the Bondi mass. Regarding the Bondi parameter $ U_0$, setting it to zero as it occasionally happens in the literature, amounts to describing a boundary fluid comoving with the natural frame of $\{u,\phi\}$.

We would like now to discuss more extensively the role of the velocity field. The redundancy of the fluid velocity in relativistic hydrodynamics, revealed originally by Eckart in the thirties, was formally introduced in \cite{Landau1987Fluid}, and more recently discussed in \cite{rezzolla2013relativistic, Kovtun:2012rj, Ciambelli:2017wou, Grozdanov:2019kge}. The argument is based on the simple fact that the energy and mass flows are indistinguishable in relativistic systems.  What matters is the energy--momentum tensor, the conserved currents (if any) and the entropy current, which do not carry any intrinsic information on a velocity field, and are therefore insensitive to it. The latter appears as an organizing artifice, which allows the 
decomposition of vectors and tensors in longitudinal and transverse components.

More concretely, a change of hydrodynamic frame is a local Lorentz boost acting on the fluid velocity $u^a$. In two dimensions, it is parametrized in the following way 
\begin{equation}
\label{locLorFinite}
\begin{pmatrix}
 u_a'\\
 \star u_a'\
\end{pmatrix}
=
 \begin{pmatrix}
 \cosh\psi && \sinh\psi\\
 \sinh\psi && \cosh\psi
 \end{pmatrix}
\begin{pmatrix}
 u_a\\
\star u_a
\end{pmatrix},
\end{equation}
where $\psi$ is any function of the boundary coordinates $u$ and $\phi$. Infinitesimally  (the suffix L stands for Lorentz)
\begin{equation}
\label{locLor}
\delta_{\text{L}}\begin{pmatrix}
 u_a\\
\star u_a\
\end{pmatrix}
=
 \psi\begin{pmatrix}
\star u_a\\
u_a
\end{pmatrix},
\end{equation}
producing 
\begin{equation}
\label{locLorThs}
\delta_{\text{L}} \Theta= \star{u}^a\pa_a\psi+\Theta^\star \psi,\quad 
\delta_{\text{L}} \Theta^\star=u^a\pa_a\psi+\Theta \psi.
\end{equation}
A change of hydrodynamic frame should leave the energy--momentum tensor invariant. Therefore \eqref{locLorFinite} should be accompanied by a change in the energy density $\varepsilon$ and the heat current $\chi$, see \eqref{energy--momentum DE}, so that the total energy--momentum tensor, i.e. $\mathcal{T}_{ab}+D_{ab}$ is unchanged.\footnote{Notice that a change of hydrodynamic frame has no effect on the geometry on which the fluid lies, in other words, it does not affect the boundary metric.  } In the same way we have decomposed the tensor $\mathcal{T}_{ab}$ into a energy density and a heat current (the trace being fixed by the geometry), we define the total energy density $\tilde{\varepsilon}$ and the total heat current $\tilde{\chi}$ in the following way -- see\cite{Campoleoni:2018ltl}
\beq
\mathcal{T}_{ab}+D_{ab}=2\ell^2\tilde{\varepsilon} u_a u_b+\tilde{\varepsilon}\tilde g_{ab}+\tau_{ab}+\ell^2 u_a \tilde{q}_b+\ell^2 u_b \tilde{q}_a,
\eeq
where $\tilde{q}_a=\tilde{\chi}\star u_a$ and $\tau_{ab}=\ell^2\tau \star u_a\star u_b$. The latter is unchanged since $D_{ab}$ is traceless. One can derive the expressions of $\tilde{\varepsilon}$ and $\tilde{\chi}$ in terms of $\varepsilon$ and $\chi$:
\begin{eqnarray}
\label{tilvarep}
\tilde\varepsilon&=&\varepsilon
+\frac{\ell^2}{8\pi G}\left(u^a\partial_a\Theta + \star u^a\partial_a\Theta^\star\right) 
- \frac{\tilde{R}}{16\pi G},
\\
\label{tilvarchi}
\tilde\chi&=& \chi -\frac{\ell^2}{4\pi G}\star u^a\partial_a \Theta
.
\end{eqnarray}
Under the transformation \eqref{locLor}, the total energy--momentum tensor is left invariant if 
\begin{equation}
\label{locLor-en-he-nc}
\delta_{\text{L}}\begin{pmatrix}
\tilde\varepsilon\\
\tilde\chi
\end{pmatrix}
=- 2\psi
\begin{pmatrix}
\tilde\chi\\
\tilde\varepsilon
\end{pmatrix}
- \psi
\begin{pmatrix}
 0\\
\tau
\end{pmatrix}, 
\end{equation}
while $\delta_{\text{L}}\tau=0$ because $\tau=\frac{\tilde{R}}{8\pi G}$. Applied to \eqref{tilvarep} and \eqref{tilvarchi}, the transformation rules \eqref{locLor}, \eqref{locLorThs} and \eqref{locLor-en-he-nc} lead to the actual energy and heat-density variations. Finally,
comparing the resulting transformations with \eqref{var3.1} and \eqref{var3.2}, one finds that they match when
\begin{equation}
\xi_{(0)}^a=0, \quad R=0 \quad \text{and}\quad  H_a=-\psi \star u_a.\label{LBb}
\end{equation}
Imposing \eqref{LBb}, the diffeomorphism \eqref{resiDE1} and \eqref{resiDE2} is a Lorentz boost on the boundary that leaves the total energy--momentum tensor unchanged. 

We would like finally to comment on the global aspects associated with the hydrodynamic-frame transformations. Actually, a change of hydrodynamic frame is a \emph{local} Lorentz boost, i.e. a \emph{gauge} transformation, and as such this invariance could possibly exhibit global drawbacks. 
Holography, as used here, provides the appropriate tool for investigating them. This was part of the agenda of \cite{Campoleoni:2018ltl}, and consists in computing the bulk conserved charges and their associated algebras, and analyzing their sensitivity to the choice of hydrodynamic frame on the boundary. The conclusion of 
\cite{Campoleoni:2018ltl} was unambiguous:  $\text{AdS}_3$ spacetimes dual to fluids with or without heat current, but with identical energy--momentum tensor, have different conserved charges, obeying different algebras.  A similar property holds in the flat limit and will be discussed in section \ref{Ricci-flat Holo}. The open question is whether the bulk diffeomorphism that we have discussed in the present section, associated with local Lorentz boosts on the boundary, is charged or not. The answer to this question would make the above claim sharper. 
 
\section{Ricci-Flat Limit}
\label{sec:Ricci Flat}

In this section, we investigate the flat limit ($\ell \to \infty$) of the results obtained in sections \ref{AdS Gauges} and \ref{Gauge Matching}, i.e. the gauge fixings, the solution spaces, the residual diffeomorphisms as well as the embedding of the Bondi gauge into the derivative expansion. This limit is well-defined in the Bondi gauge and in the derivative expansion, while the Fefferman--Graham gauge blows up due to the component $g_{\rho \rho} = \frac{\ell^2}{\rho^2}$. As we will explain shortly, a subtle point is here the necessity of prescribing the large-$\ell$ behavior for the data in the derivative expansion before taking the limit.

\subsection{Bondi Gauge} 
\label{Bondi Gauge subsection}

The definition of the Bondi gauge (\ref{Bondi metric},\ref{Bondi gauge fixing}) and the associated off-shell residual gauge diffeomorphisms (\ref{resB1}-\ref{resB3}) are valid irrespective of the value of 
$\ell$. In the following, we investigate the solution space and the on-shell residual gauge diffeomorphism in the Ricci-flat case, i.e. for infinite $\ell$. This extends previous analysis to generic boundary data, i.e. asymptotically locally flat spacetimes.

\subsubsection*{Solution space}

In three dimensions, the full solution space in Bondi gauge for vanishing cosmological constant can be readily obtained by taking the flat limit of the solution space obtained in section \ref{Solution space} for non-vanishing cosmological constant. In practice, we take $\frac{1}{\ell} \to 0$ in the equations. The equation $G_{rr} = 0$ gives
\begin{equation}
\beta = \beta_0 (u, \phi)
\end{equation} Solving $G_{r\phi} = 0$ leads to
\begin{equation}
U= U_0(u, \phi) + \frac{1}{r}  2 e^{2\beta_0} e^{-2 \varphi} \partial_\phi \beta_0  - \frac{1}{r^2} e^{2\beta_0} e^{-2 \varphi} N(u, \phi) .
\end{equation} Solving $G_{u r} = 0$ gives
\begin{equation}
\frac{V}{r} =  - 2 r (\partial_u \varphi + D_\phi U_0 ) + M (u, \phi) + \frac{1}{r} 4 e^{2 \beta_0} e^{-2 \varphi} N \partial_\phi \beta_0 - \frac{1}{r^2} e^{2 \beta_0} e^{-2 \varphi} N^2 ,
\end{equation} where $D_\phi U_0 = \partial_\phi U_0 + \partial_\phi \varphi U_0$. Taking into account the previous results, the Einstein equation $G_{\phi\phi} =0$ is satisfied at all orders. Finally, we solve the Einstein equations giving the time evolution constraints on $M$ and $N$. The equation $G_{u\phi} =0$ gives
\beqn
(\partial_u + \partial_u \varphi ) N &=& \left(\frac{1}{2} \partial_\phi + \partial_\phi \beta_0 \right) M -2 N \partial_\phi U_0 - U_0 (\partial_\phi N + N \partial_\phi \varphi )\nonumber \\
&&+ 4 e^{2 \beta_0} e^{-2 \varphi} [2 (\partial_\phi \beta_0)^3 - (\partial_\phi \varphi) (\partial_\phi \beta_0)^2 + (\partial_\phi \beta_0 ) (\partial_\phi^2 \beta_0) ] ,
\eeqn 
whereas $G_{uu}= 0$ results in 
\beqn
\partial_u M &=&(- 2 \partial_u \varphi + 2 \partial_u \beta_0-2\pa_\phi U_0+U_0 2\partial_\phi \beta_0- U_0 2\partial_\phi \varphi-U_0\partial_\phi)  M-2 e^{2 \beta_0} e^{-2 \varphi} \{\partial_\phi U_0 [8 (\partial_\phi \beta_0)^2+ (\partial_\phi \varphi)^2  \nonumber\\
&& - 4 \partial_\phi \beta_0 \partial_\phi \varphi + 4 \partial_\phi^2 \beta_0 - 2 \partial_\phi^2 \varphi ] -\partial_\phi^3 U_0 + U_0 [ \partial_\phi \beta_0 (8 \partial_\phi^2 \beta_0 - 2 \partial_\phi^2 \varphi ) + 2 \partial_\phi^3 \beta_0 - \partial_\phi^3 \varphi+ \partial_\phi \varphi (- 2 \partial_\phi^2 \beta_0 \nonumber\\
&& + \partial_\phi^2 \varphi )  ]  + 2 \partial_u \partial_\phi \beta_0 (4 \partial_\phi \beta_0 - \partial_\phi \varphi ) + \partial_u \partial_\phi \varphi (-2 \partial_\phi \beta_0 + \partial_\phi \varphi ) + 2 \partial_u \partial_\phi^2 \beta_0 - \partial_u \partial_\phi^2 \varphi \}.
\eeqn
 The solution space is thus characterized by five arbitrary functions of $(u, \phi)$, given by $\beta_0$, $U_0$, $M$, $N$, $\varphi$, with two dynamical constraints given by the time evolution equations of $M$ and $N$. 

\subsubsection*{Variation of the solution space}

By a similar procedure, the on-shell residual gauge diffeomorphisms and the variations of the solution space are obtained by taking $\ell \to \infty$ in the expressions (\ref{resBON1}-\ref{resBON3}) and (\ref{var1}-\ref{var2}), respectively. 
The on-shell residual gauge diffeomorphisms are given by
\beqn
\xi^u &=& f,\label{resBON1flat1}\\
\xi^\phi &=& Y - \frac{1}{r} \partial_\phi f\, e^{2\beta_0-2\varphi} ,\label{resBON2flat2}\\
\xi^r &=& - r [ \partial_\phi Y - \omega - U_0 \partial_\phi f + Y \partial_\phi \varphi + f \partial_u \varphi ] \nonumber\\
&&+ e^{2\beta_0 - 2 \varphi} (\partial_\phi^2 f - \partial_\phi f \partial_\phi \varphi + 4 \partial_\phi f \partial_\phi \beta_0) - \frac{1}{r} e^{2\beta_0 - 2 \varphi}  \partial_\phi f\, N .\label{resBON3flat3}
\eeqn
Under these residual gauge diffeomorphisms, the unconstrained part of the solution space transforms as
\beqn
\delta_\xi \varphi &=& \omega, \\
\delta_\xi \beta_0 &=& (f \partial_u + Y \partial_\phi)\beta_0 + \left(\frac{1}{2}\partial_u - \frac{1}{2} \partial_u \varphi + U_0 \partial_\phi \right) f - \frac{1}{2}(\partial_\phi Y + Y \partial_\phi \varphi - \omega ), \\
\delta_\xi U_0 &=& (f \partial_u + Y \partial_\phi - \partial_\phi Y ) U_0 -\partial_u Y + U_0 (\partial_u f +  U_0 \partial_\phi f) ,
\eeqn 
while the constrained part transforms as
\beqn
\delta_\xi N &=& (f \partial_u + Y \partial_\phi + 2 \partial_\phi Y +f \partial_u \varphi+  Y \partial_\phi \varphi -\omega-2 U_0 \partial_\phi f )N + M \partial_\phi f -e^{2\beta_0-2\varphi}[3\partial_\phi^2 f (2\partial_\phi \beta_0 - \partial_\phi \varphi)\nonumber \\
&& + \partial_\phi f( 4 (\partial_\phi \beta_0)^2 - 8 \partial_\phi \beta_0 \partial_\phi \varphi + 2 (\partial_\phi \varphi)^2 + 2 \partial_\phi^2 \beta_0 -\partial_\phi^2 \varphi)+ \partial_\phi^3 f  ] , \\
\delta_\xi M &=&(\pa_u f+f \pa_u\varphi+\pa_\phi Y+Y\pa_\phi\varphi-\omega)M-2e^{2\beta_0-2\varphi}\Big[2\pa_\phi^2f\pa_u\beta_0+4\pa_u\pa_\phi f\pa_\phi\beta_0+\pa_u\pa_\phi^2 f \nonumber \\
&&+\pa_\phi^2f\pa_\phi U_0+8\pa_\phi^2 f\pa_\phi \beta_0 U_0+\pa_\phi f\Big((4\pa_\phi\beta_0-\pa_\phi\varphi)(2\pa_u \beta_0-\pa_u \varphi)+4\pa_u\pa_\phi\beta_0+\pa_\phi U_0(8\pa_\phi\beta_0-2\pa_\phi\varphi)\nonumber\\&&-\pa_\phi^2 U_0-2\pa_u\pa_\phi\varphi
+U_0(-4\pa_\phi\beta_0\pa_\phi\varphi+8(\pa_\phi\beta_0)^2+4\pa_\phi^2\beta_0+(\pa_\phi\varphi)^2-2\pa_\phi^2\varphi)\Big)-2\pa_\phi^2 f U_0\pa_\phi\varphi+\pa_\phi^3 fU_0\nonumber\\&&-\pa_u\pa_\phi f\pa_\phi\varphi-\pa_\phi^2 f\pa_u\varphi-2f\pa_\phi\beta_0\pa_u\pa_\phi\varphi+2\pa_\phi\beta_0\pa_\phi\omega-2\pa_\phi\beta_0\pa_\phi Y\pa_\phi\varphi-2\pa_\phi\beta_0\pa_\phi^2 Y-2\pa_\phi \beta_0 Y\pa_\phi^2\varphi\Big]\nonumber\\&&+f\pa_u M+\pa_\phi M Y .
\eeqn

We can impose an equivalent of the Brown--Henneaux boundary condition in flat space by asking 
\begin{equation}
\varphi=0, \quad \beta_0=0,\quad U_0=0.
\end{equation} 
The first one fixes $\omega$ to be zero, the second and third ones lead to two equations on $f$ and $Y$
\begin{equation}
\partial_u f =\partial_\phi Y, \quad \partial_u Y=0.
\end{equation}
The solutions to these two equations describe the so-called BMS$_3$ algebra. The asymptotic Killing is then uniquely specified by a couple $(f,Y)$ belonging to BMS$_3$.

\subsection{Carrollian Fluid/Gravity Derivative Expansion}
\label{subsec:Deriative expansion}

We study here the flat limit of the derivative expansion. Following \cite{CMPR}, this is performed without calling for a specific parametrization of the boundary geometry. The salient features of this analysis are (\romannumeral1) the appearance of a two-dimensional boundary for asymptotically flat spacetimes, located at null infinity and equipped with a Carrollian structure; (\romannumeral2) the emergence of a Carrollian fluid, obeying Carrollian hydrodynamics, and carrying the degrees of freedom dual to the flat bulk spacetime. 

\subsubsection*{Flat limit of the bulk metric}

The behavior of the various relativistic boundary tensors in the $\ell\to \infty$ limit is not universal. Nonetheless, some properties are, such as the fact that the velocity is becoming null-like and the boundary metric degenerates \cite{Campoleoni:2018ltl}. Inspired by the explicit results in this reference, and in line with \cite{CMPR},\footnote{In comparing with this reference, we should trade 
$\alpha$ for $\zeta$ ($\chi_\pi$ in \cite{Campoleoni:2018ltl}) and  $\Xi$ for $\theta$.} we set:
\begin{align}
\mu_a &= \lim_{\ell \to \infty} \ell^2 u_a ,\\
v^a &= \lim_{\ell \to \infty} u^a, \\
 v_\star^a &= \lim_{\ell \to \infty} \ell \star u^a ,\\
\mu^\star_a &= \lim_{\ell \to \infty} \ell \star u_a, \\
\alpha &= \lim_{\ell\to\infty} \ell \chi ,\\
\epsilon &= \lim_{\ell\to\infty} \varepsilon .
\end{align} 
The boundary metric reads
\beq
\tilde g_{ab}=\ell^2(-u_a u_b+\star u_a \star u_b).
\eeq
Hence, we gather in the $\ell \to \infty$ limit
\beq
\star\mu_a\star \mu_b=\lim_{\ell\to\infty}\tilde g_{ab}.
\eeq
The flat limit leads therefore to a degenerate boundary metric. This is the well-known ultra-relativistic limit, named Carrollian in \cite{Levy}, and emerging generally as the geometry of null hypersurfaces (here null infinity) \cite{Duval:2014uoa, Duval:2014uva, Duval:2014lpa, Bekaert:2014bwa, Bekaert:2015xua, Hartong:2015xda, Figueroa-OFarrill:2018ilb, Morand:2018tke, Ciambelli:2019lap, Bergshoeff:2014jla, card, ba, CM1} (see also section \ref{Ricci-flat Holo}).
We can also define a density
\begin{equation}
\mathcal{D} = \lim_{\ell \to \infty} \ell \sqrt{-\tilde{g}} = |\varepsilon^{ab} \mu_a \mu^\star_b |,
\end{equation}
which allows expressing the composed quantities
\begin{align}
\Xi &= \lim_{\ell \to \infty} \Theta = \partial_av^a+v^a\pa_a\ln {\cal D},\\
\Xi^\star &=   \lim_{\ell \to \infty} \ell \Theta^\star = \partial_a  v^a_\star + v^a_\star\pa_a \ln \mathcal{D},
\end{align}
and
\beq\label{calA}
{\cal A}_a= \lim_{\ell \to \infty} A_a =\mu^\star_a \Xi^\star-\mu_a \Xi.
\eeq

From the functions $(u_a, \tilde{g}_{ab}, \varepsilon, \chi)$ parametrizing the derivative expansion in the asymptotically locally AdS$_3$ case, where $u^a$ is normalized as $u^a \tilde{g}_{ab} u^b = -\frac{1}{\ell^2}$, and $\varepsilon$ and $\chi$ satisfy time evolution equations, we get the following set of functions in the Ricci-flat limit: $(\mu_a , v^a , v^a_\star, \mu_a^\star , \alpha, \epsilon)$. Let us now determine the constrains between these objects, induced from AdS.  Firstly, defining the tensors
\begin{equation}
\mathcal{D}_{ab} = \mathcal{D} \varepsilon_{ab}, \qquad \mathcal{D}^{ab} = -\mathcal{D}^{-1} \varepsilon^{ab},
\end{equation}
the duality relations give
\begin{align}
\ell \star u_a = \ell \eta_{ab} u^b \quad &\xrightarrow[\ell \to \infty]{} \quad \mu_a^\star = \mathcal{D}_{ab} v^b,\label{H1} \\
\ell \star u^a = \ell  \eta^{ab} u_b \quad &\xrightarrow[\ell \to \infty]{} \quad v^a_\star = \mathcal{D}^{ab} \mu_b ,\label{H2} \\
\ell^2 u_a =  \ell^2 \eta_{ab} \star u^b \quad &\xrightarrow[\ell \to \infty]{} \quad \mu_a = \mathcal{D}_{ab} v^b_\star, \label{H3}\\
u^a =  \eta^{ab} \star u_b  \quad &\xrightarrow[\ell \to \infty]{} \quad v^a =  \mathcal{D}^{ab} \mu_b^\star.\label{H4}
\end{align} 
Out of eight equations, we have four independent constraints. This can be seen imposing for instance $v^a_\star = \mathcal{D}^{ab} \mu_b$ and $v^a =  \mathcal{D}^{ab} \mu_b^\star$. These four equations automatically imply the other four.

We now show that every other constraint arising in the limit does not imply new independent equations. First, projecting the previous expressions on the vectors $v^a$, $v^a_\star$ and the forms $\mu_a$, $\mu_a^\star$, we obtain:
\beqn
 \mu_a^\star v^a = 0, &\quad & \mu_a^\star v^a_\star=v^a_\star {\cal D}_{ab}v^b=1,\\
\mu_a v^a_\star = 0, &\quad & \mu_a^\star v^a_\star=\mu^\star_a {\cal D}^{ab}\mu_b=1,\\
 \mu_a v^a_\star = 0, &\quad & \mu_a v^a=v^a {\cal D}_{ab}v^b_\star=-1,\\
\mu^\star_a v^a = 0, &\quad  & \mu_a v^a=\mu_a {\cal D}^{ab}\mu^\star_b=-1.
\eeqn
The normalization of $u^a$ and its different equivalent avatars AdS lead to
\begin{align}
\tilde{g}_{ab} u^a u^b = - \frac{1}{\ell^2} \quad &\xrightarrow[\ell \to \infty]{} \quad  \mu^\star_a v^a = 0, \\
\ell^2 \tilde{g}^{ab} u_a u_b = -1 \quad &\xrightarrow[\ell \to \infty]{} \quad \mu_a v^a = -1 ,\\
\ell^2 \tilde{g}_{ab} \star u_a \star u_b = 1 \quad &\xrightarrow[\ell \to \infty]{} \quad \mu_a^\star v^a_\star = 1 ,\\
\tilde{g}^{ab} \star u_a \star u_b = \frac{1}{\ell^2} \quad &\xrightarrow[\ell \to \infty]{} \quad \mu^\star_a v^a = 0,
\end{align} and
\begin{align}
\ell \tilde{g}_{ab} u^a \star u^b = 0 \quad &\xrightarrow[\ell \to \infty]{} \quad (\mu^\star_a v^a ) (\mu_b^\star v^b_\star) = 0, \\
\ell  \tilde{g}^{ab} u_a \star u_b = 0 \quad &\xrightarrow[\ell \to \infty]{} \quad (\mu_a v^a) (\mu_b^\star v^b ) = 0.
\end{align}
Moreover, the relations between the vectors and their co-vectors yield
\begin{align}
u_a = \tilde{g}_{ab} u^b \quad &\xrightarrow[\ell \to \infty]{} \quad  \mu_a^\star v^a = 0 ,\\
u^a = \tilde{g}^{ab} u_b \quad &\xrightarrow[\ell \to \infty]{} \quad \mu_a v^a = -1, \\
\ell \star u_a = \ell \tilde{g}_{ab} \star u^b \quad &\xrightarrow[\ell \to \infty]{} \quad \mu^\star_a v^a_\star = 1 ,\\
\frac{1}{\ell}\star u^a = \frac{1}{\ell}\tilde{g}^{ab} \star u_b \quad &\xrightarrow[\ell \to \infty]{} \quad  \mu_a^\star v^a = 0.
\end{align}  
These equations are indeed all consequences of (\ref{H1}-\ref{H4}), leading to no further constraints. Notice that, using the dual relations, we have
\begin{equation}
\Xi = \mathcal{D}^{cd} \partial_c \mu^\star_d, \qquad \Xi^\star = \mathcal{D}^{cd} \partial_c \mu_d. 
\label{XiDef}
\end{equation}

In summary, we have $10$ functions in $(\mu_a , v^a , v^a_\star, \mu_a^\star ,  \alpha, \epsilon)$, with $4$ independent constraints, which makes a total of $6$ independent functions. There is some freedom to decide which functions we keep free and which functions we fix through the constraints. We propose an interesting choice of parametrization inspired by the solution space in AdS, that makes the interpretation easier in the flat limit. We choose as free data $(\mu_a , \mu_a^\star , \alpha, \epsilon)$ . It is straightforward to see that this is actually a minimum choice of functions to re-construct the other functions $(v^a, v^a_\star)$ through $v^a =  \mathcal{D}^{ab} \mu_b^\star$ and $v^a_\star = \mathcal{D}^{ab} \mu_b$, and to satisfy the remaining constraints.  As discussed in section \ref{Gauge matching flat}, this specific parametrization is well-adapted to relate the derivative expansion with the Bondi gauge.

Under the above assumptions of behavior in $\ell$, the derivative expansion \eqref{DE} admits a well-defined flat limit. We obtain
\begin{equation}
\D s^2_{\text{Flat}} = \lim_{\ell\to \infty} \D s^2_{\text{AdS}} = 2 \mu_a \D x^a (\D r+r{\cal A}_b \D x^b )+r^2 \mu^\star_a \mu^\star_b \D x^a \D x^b +8\pi G \mu_a \left(\epsilon \mu_b + \alpha \mu^\star_b \right) \D x^a \D x^b.
\label{flat DE}
\end{equation} 
Furthermore, the limit $\ell\to\infty$ of (\ref{eom1}-\ref{eom2}) gives the Ricci-flat Einstein equations for the metric \eqref{flat DE}
\beqn
(v^a \partial_a  +2 \Xi ) \epsilon - \frac{1}{4 \pi G} (v^a_\star \partial_a + 2 \Xi^\star ) (v^a_\star \partial_a \Xi - v^a \partial_a \Xi^\star ) &=& 0, \\
(v^a_\star \partial_a  +2 \Xi^\star ) \epsilon + (v^a \partial_a +2 \Xi) \alpha &=&0.\label{flat2}
\eeqn
The bulk metric and the conservation equations follow a similar motive compared to  anti de Sitter with $(\varepsilon, \chi,u_a,\star u_a)$ traded for $(\epsilon,\alpha,\mu_a,\mu^\star_a)$.  As already mentioned, in the original AdS case, the bulk was locally $\text{AdS}_3$, the two-dimensional boundary pseudo-Riemannian, and the evolution equations were relativistic. In the case at hand, the bulk is three-dimensional locally Minkowski, whereas the  boundary and the evolution equations are Carrollian.  

\subsubsection*{On-shell residual gauge diffeomorphisms}

We follow the same procedure as in the AdS case in order to determine the residual gauge diffeomorphisms of the Ricci flat derivative expansion \eqref{flat DE}. We start from the following ansatz:
\begin{equation}
\xi^a = \xi^a_{(0)} + \frac{1}{r} \xi^a_{(1)} , \quad \xi^r = r R + V + \frac{1}{r} W ,
\label{ansatz DE v2}
\end{equation} where $\xi^a_{(0)}= (F, \mathcal{Y})$, $\xi^a_{(1)} = (L, Z)$, $R$, $V$ and $W$ are functions of $x^a$. As for anti de Sitter, the form of these vectors is motivated by the expression of the residual gauge diffeomorphisms in Bondi gauge (\ref{resBON1flat1}-\ref{resBON3flat3}) and the gauge matching that will be established in section \ref{Gauge matching flat}. The condition $\mathcal{L}_\xi g_{rr} = 0$ leads to
\begin{equation}
\mu_a \xi^{a}_{(1)} = 0  \quad \Longleftrightarrow \quad L = - \frac{\mu_\phi}{\mu_u} Z . 
\label{constraint xi1}
\end{equation} Now, let us consider the variations 
\begin{equation}
\delta_\xi \mu_a = h_a, \qquad \delta_\xi \mu_a^\star = h^\star_a,
\end{equation} where we introduced the parameters $h_a$ and $h^\star_a$ as functions of $x^a$. These functions are not all independent. Indeed, ${\cal L}_\xi g_{ab}|_{\text{order }r^2}=\delta_\xi (\mu^\star_a\mu^\star_b)$ gives
\begin{equation}
h^\star_a = \mathcal{L}_{\xi_{(0)}} \mu^\star_a +  R  \mu^\star_a. 
\label{h star equation}
\end{equation}
From $\delta_\xi \mu_a = \mathcal{L}_\xi g_{ra}|_{\text{order }1}$, we get the two equations 
\begin{align}
&(\mathcal{L}_{\xi_{(0)}} \mu_a + R \mu_a - h_a ) \mathcal{D}^{ab} \mu^\star_b = 0, \label{first rel} \\
&(\mathcal{L}_{\xi_{(0)}} \mu_a - h_a ) \mathcal{D}^{ab} \mu_b -  \xi_{(1)}^a  \mu_a^\star = 0 . \label{second rel}  
\end{align}  Using \eqref{constraint xi1}, the second equation \eqref{second rel} leads to 
\begin{equation}
\xi_{(1)}^a = [(\mathcal{L}_{\xi_{(0)}} \mu_b - h_b )\mathcal{D}^{bc} \mu_c] v^a_\star . 
\label{expression xi 1 flat}
\end{equation} The first relation \eqref{first rel} allows us to derive\footnote{The function $\mathcal{D}$ is a density, therefore its Lie derivative has an additional divergence term: $\mathcal{L}_{\xi^{(0)}}\ln\mathcal{D}=\xi_{(0)}^a\partial_a \ln\mathcal{D}+\partial_a\xi_{(0)}^a$.}
\begin{equation}
\delta_\xi \ln \mathcal{D}= {\cal L}_{\xi^{(0)}}\ln \mathcal{D}  + 2 R  ,
\end{equation}
from which one can choose to express one parameter in terms of the other. We decide to express $h_u$ as
\beq
h_u={\mu^\star_u h_\phi\over \mu^\star_\phi}+{{\cal D}\over \mu_\phi^\star} \Big(R+{\cal L}_{\xi^{(0)}}\ln{\cal D}-v^a_\star {\cal L}_{\xi^{(0)}}\mu^\star_a\Big).\label{hu}
\eeq

We then require $\mathcal{L}_\xi g_{ra}|_{\text{order }1/r^2} = 0$ and find 
\begin{equation}
W = - 4 \pi G \alpha \xi^a_{(1)} \mu_a^\star ,
\end{equation} while $\mathcal{L}_\xi g_{ra}|_{\text{order }1/r} = \mathcal{L}_{\xi_{(1)}} \mu_a - \mu_a \xi_{(1)}^b \mu^\star_b \Xi^\star = 0$ does not give further constraint. We turn our attention to the equations $\mathcal{L}_\xi g_{ab}|_{\text{order }r} = \delta_\xi (\mu_a {\cal A}_b+\mu_b {\cal A}_a)$, which read
\beq\label{Or}
\delta_\xi (\mu_a {\cal A}_b+\mu_b {\cal A}_a)={\cal L}_{\xi^{(0)}}(\mu_a {\cal A}_b+\mu_b {\cal A}_a)+{\cal L}_{\xi^{(1)}}(\mu_a^\star \mu_b^\star)+R(\mu_a {\cal A}_b+\mu_b {\cal A}_a)+2 V\mu^\star_a\mu^\star_b+\pa_a R \mu_b+\pa_b R\mu_a.
\eeq
Noticing that $v^a{\cal A}_a=\Xi$ and $v^a_\star {\cal A}_a=\Xi^\star$, imposing \eqref{first rel} and projecting \eqref{Or} on $v^a$ and $v^a_\star$, we gather\footnote{We use identities like $v^a v^b_\star-v^b v^a_\star={\cal D}^{ab}$, $v_\star^a\mu_b^\star-v^a\mu_b=\delta^{a}_{b}$, $\mu_a \mu_b^\star-\mu_b\mu_a^\star={\cal D}_{ab}$ and $\D {\cal D}_{ab}={\cal D}_{ab} \D \ln {\cal D}$.}
\beqn
v^a_\star v^b_\star \ \eqref{Or} & \Leftrightarrow & V=-\pa_a \xi^a_{(1)}-\xi^a_{(1)}\pa_a \ln {\cal D}\\
v^a v^b_\star \ \eqref{Or} & \Leftrightarrow & v^b_\star \delta_\xi {\cal A}_b=v^b_\star \pa_b (v^a (\mathcal{L}_{\xi_{(0)}} \mu_a - h_a ))-v^b\pa_b (v^a_\star (\mathcal{L}_{\xi_{(0)}} \mu_a - h_a ))+v^b_\star {\cal L}_{\xi^{(0)}}{\cal A}_b\label{dA1}\\
v^a v^b \ \eqref{Or} & \Leftrightarrow & v^b \delta_\xi {\cal A}_b=v^a\pa_a(v^b(\mathcal{L}_{\xi_{(0)}} \mu_b - h_b ))+v^b {\cal L}_{\xi^{(0)}}{\cal A}_b.\label{dA2}
\eeqn
The first equation constraints $V$, whereas, using the explicit expression for ${\cal A}_a$, \eqref{calA}, one can show that \eqref{dA1} and \eqref{dA2} are identities, which therefore do not infer additional constraints. There are two other sets of equations we need to solve, namely $\mathcal{L}_\xi g_{ab}|_{\text{order }1} = \delta_\xi(8\pi G \epsilon \mu_a\mu_b+4\pi G\alpha (\mu_a\mu^\star_b+\mu_b\mu^\star_a))$ and $\mathcal{L}_\xi g_{ab}|_{\text{order }1/r} = 0$. The latter gives two equations: one is automatically satisfied and the other is proportional to \eqref{flat2}, thus also satisfied on-shell. The former encodes two independent equations, which give the variations of $\epsilon$ and $\alpha$, as outlined in the next section.

Putting everything together, the family of residual gauge diffeomorphisms considered in \eqref{ansatz DE v2} is explicitly given by 
\beqn
\xi^a &=& \xi^a_{(0)} + \frac{1}{r} \xi^a_{(1)},\label{xi flat 1} \\
\xi^r &=& r R -(\pa_a +\pa_a \ln {\cal D}) \xi^a_{(1)} - \frac{4}{r} \pi G \alpha \xi^a_{(1)} \mu_a^\star \label{xi flat 2},
\eeqn 
where $\xi^a_{(1)}$ is written in \eqref{expression xi 1 flat}. These diffeomorphisms are parametrized by four arbitrary functions of $(u, \phi)$ given by $\xi^a_{(0)} = (F, \mathcal{Y})$, $R$ and $h_\phi$. Notice that there is an additional parameter $h_\phi$ compared to the residual gauge diffeomorphisms of the Bondi gauge (\ref{resBON1flat1}-\ref{resBON3flat3}). 
In particular, as discussed in section \ref{Gauge matching flat}, we can at any moment set consistently $\mu_\phi = 0$ and $h_\phi = 0$, and obtain the Bondi residual gauge diffeomorphisms. 

An alternative way to obtain (\ref{xi flat 1}-\ref{xi flat 2}) is to take directly the flat limit on the residual gauge diffeomorphisms (\ref{resiDE1}-\ref{resiDE2}) of the derivative expansion in AdS, with the appropriate falloffs in $\ell$ on the parameters:
\begin{align}
\xi^a_{(0), \text{Flat}} &= \lim_{\ell \to \infty} \xi^a_{(0), \text{AdS}}, \\
R_{\text{Flat}} &= \lim_{\ell \to \infty} R_{\text{AdS}}, \\
h_\phi &=  \lim_{\ell \to \infty} \ell^2 H_\phi .
\end{align} 
Doing so, we get exactly (\ref{xi flat 1}-\ref{xi flat 2}) in the flat limit.

\subsubsection*{Variation of the solution space}

Under infinitesimal diffeomorphisms generated by (\ref{xi flat 1}-\ref{xi flat 2}), the first part of the solution space transforms as
\beqn
\delta_\xi \mu_a^\star &=& \mathcal{L}_{\xi_{(0)}} \mu^\star_a +  R  \mu^\star_a \\
\delta_\xi \mu_a &=& h_a,
\eeqn 
with $h_u$ given by \eqref{hu}. By analogy with the anti-de Sitter situation, this is consistent with the fact that $\mu_\phi$ will be the additional parameter that the derivative expansion possesses compared to the Bondi gauge; it is the Ricci-flat equivalent of the parameter $u_\phi$. In section \ref{Ricci-flat Holo}, the parameter $\mu_a$ (or equivalently $v^a_\star$) will be interpreted as the velocity of an ultra-relativistic fluid. The second part of the solutions space transforms as
\beqn
\delta_\xi \alpha &=& -2\alpha R+{\cal L}_{\xi^{(0)}}\alpha+2\epsilon \xi^a_{(1)}\mu^\star_a+{1\over 4\pi G}\left(v^a_\star {\cal L}_{\xi^{(1)}}{\cal A}_a-\Xi^\star v^a{\cal L}_{\xi^{(1)}}\mu_a+V\Xi^\star+v^b_\star \pa_b V\right) ,\\
\delta_\xi \epsilon &=& -2\epsilon R+{\cal L}_{\xi^{(0)}}\epsilon-{1\over 4\pi G}\left(v^a {\cal L}_{\xi^{(1)}}{\cal A}_a-\Xi v^a{\cal L}_{\xi^{(1)}}\mu_a+V\Xi+v^b \pa_b V\right).
\label{Transfo-alpha-epsilon}
\eeqn

\subsection{Gauge Matching}
\label{Gauge matching flat}

The Bondi gauge for locally flat spacetimes can be seen as a sub-gauge of the flat derivative expansion. We give here the dictionary relating the solution spaces and the parameters of the residual gauge diffeomorphisms. 

\subsubsection*{Bondi gauge, derivative expansion and solution spaces matching}

Imposing $\mu_\phi = 0$ in the Ricci-Flat derivative expansion \eqref{flat DE}, we are precisely in the Bondi gauge. The two solution spaces are then identified as\footnote{In the computations above, we assumed ${\cal D}=|\epsilon^{ab}\mu_a\mu^\star_b|=\epsilon^{ab}\mu_a\mu^\star_b$, i.e. $\epsilon^{ab}\mu_a\mu^\star_b$ being positive. This fixes the relative signs and ambiguity in the comparison with the Bondi gauge.}
\beqn
\mu_u &=& - e^{2\beta_0}, \label{IndentBDE 1 v2} \\
\mu^\star_u &=& e^{\varphi} U_0 , \\
\mu^\star_\phi &=& -e^\varphi \\
\alpha &=& \frac{1}{4\pi G} e^{-\varphi} N ,\\
\epsilon &=& \frac{1}{8\pi G} (e^{-2 \beta_0}M + 4 e^{-2 \varphi} (\partial_\phi \beta_0)^2 )  . \label{IndentBDE 2 v2} 
\eeqn
Again, the Carrollian-fluid heat and energy densities are simply related to the angular momentum aspect and the Bondi mass. 

As for anti de Sitter, we could go one step further by considering the diffeomorphism that maps the Bondi gauge to the derivarive expansion with arbitrary parameter $\mu_\phi$. In particular, the flat limit of the diffeomorphism considered in (\ref{Bondi-DE for uphi non 0 1}-\ref{Bondi-DE for uphi non 0 1}) would map the Bondi gauge to the derivative expansion with arbitrary $\mu_\phi$. As discussed above, the formula describing the solution space matchings through the diffeomorphism are lengthy and not illuminating. 

\subsubsection*{Residual gauge parameters matching}

The parameters of the residual gauge diffeomorphisms of the Ricci-flat derivative expansion are related to those of the Bondi gauge as
\beqn
F &=& f, \\
\mathcal{Y} &=& Y, \\
R &=&  - \partial_\phi Y + \omega + U_0 \partial_\phi f - Y \partial_\phi \varphi - f \partial_u \varphi ,
\eeqn and $h_\phi$ has no equivalent in Bondi gauge, since it is associated to shifts of $\mu_\phi$.

\subsection{The Holographic Fluid Dual to Ricci-Flat Spacetimes}
\label{Ricci-flat Holo}

We would like to make some further comments regarding the boundary structure that emerges in the flat limit of the derivative expansion. We recall that in the AdS case, the bulk geometry described by the line element \eqref{DE} is holographically dual to a relativistic fluid. The latter is caracterised by its energy density $\varepsilon$, its heat current $\chi$,\footnote{We recall that the heat current is transverse to the fluid velocity, therefore in two dimensions, thanks to Hodge duality and without loss of generality, we can describe the heat current in terms of a scalar $\chi$.} its velocity $u^a$ and the boundary metric of the pseudo-Riemannian spacetime on which the fluid flows. The dynamics of the fluid is then captured by the conservation of the corresponding energy--momentum tensor \eqref{energy--momentum DE}. 

What happens on the boundary when the flat limit is taken in the bulk, has been discussed from several perspectives, all converging towards the emergence of Carrollian geometry and Carrollian hydrodynamics. This has led to the study of Carrollian fluids precisely in this spirit \cite{Ciambelli:2018xat}, and to their use for unravelling flat holography from the fluid/gravity side \cite{Ciambelli:2018wre, Campoleoni:2018ltl,CMPR}. Let us review how the logic goes, in the present three-dimensional paradigm. 

In locally flat spacetimes, as those found in the present work, we also have a set of boundary data, $\mu_a$, $\mu^\star_a$, $\alpha$ and $\epsilon$, and they also satisfy conservation equations \eqref{flat2}. Remember that the boundary metric of AdS has the following behavior in the flat limit:
\begin{equation}
\tilde{g}_{ab}=\ell^2(-u_au_b+\star u_a \star u_b)\underset{\ell\rightarrow\infty}{\rightarrow}\mu^\star_a\mu^\star_b\equiv h_{ab}.
\end{equation}
This metric possesses a kernel generated by the vector field $v^a$: 
\begin{equation}
\mu^\star_av^a=0.
\end{equation} 
From the bulk side this is the signature that the boundary becomes null and that the induced metric is  degenerate. Physically, this is an ultra-relativistic limit with $\ell^{-1}$ playing the role of the boundary velocity of light: the zero-$c$ limit ($\ell\rightarrow\infty$) of a relativistic metric possesses a degenerate direction that coincides with the time. The right geometrical structure that emerges and replaces the pseudo-Riemannian metric is called a Carroll manifold --  see e.g. \cite{Duval:2014uoa} for a rigorous mathematical definition. It is formulated in terms of a degenerate metric, here $h_{ab}$, and a vector field that belongs to the kernel of the metric, here $v^a$. 

Regarding the physical degrees of freedom and their description, we follow the paradigm of ordinary non-relativistic fluids. Those emerge in the Galilean limit of relativistic fluids, when $c\rightarrow\infty$. In the same manner, one can define Carrollian fluids that flow on Carroll manifolds and obey suitable conservation equations. In \cite{Ciambelli:2018xat}, the authors give a comprehensive description of Carrollian fluids and of their dynamics inherited from a controlled ultra-relativistic limit (see also \cite{Campoleoni:2018ltl} for the specific description of the two-dimensional case and \cite{Donnay:2019jiz} for an application to near-horizon physics). 

In the situation at hand, the Carrollian fluid is characterized by its energy density $\epsilon$, its Carrollian equivalent of the the heat current $\alpha$, its velocity $v^a_\star$ and it flows on a Carroll manifold $(h_{ab}, v^a)$. The conservation equations satisfied by this fluid cannot be written as the conservation of an energy--momentum tensor, simply because there is no canonical Levi--Civita connection associated with a Carroll manifold. A consistent way to obtain them is what we have done here, i.e. write explicitly the relativistic fluid equations (\ref{eom1}-\ref{eom2}) and compute their ultra-relativistic limit, which on the boundary is equivalent to a $\ell\rightarrow\infty$ limit. We recall the result 
\beqn
(v^a \partial_a  +2 \Xi ) \epsilon - \frac{1}{4 \pi G} (v^a_\star \partial_a + 2 \Xi^\star ) (v^a_\star \partial_a \Xi - v^a \partial_a \Xi^\star ) &=& 0, \\
(v^a_\star \partial_a  +2 \Xi^\star ) \epsilon + (v^a \partial_a +2 \Xi) \alpha &=&0,
\eeqn
where $\Xi$ and $\Xi^\star$ are simply first derivatives of $\mu^\star_a$ and $v_\star^a$. The lesson to be learned here is that, in three bulk dimensions, there is a non-trivial flat-space limit of fluid/gravity correspondence, and the dual to an asymptotically flat spacetime is a Carrollian fluid. Things go the same way in higher dimensions (see \cite{Ciambelli:2018wre} for a study of the four-dimensional case).

It is remarkable that the notion of hydrodynamic frame, persists for Carrollian fluids. In the relativistic case we could use a local Lorentz boost to implement a field redefinition that leaves the fluid energy--momentum tensor invariant (therefore also its equations of motion). A similar transformation exists for Carrollian fluids, now carried by a local Carrollian boost. The latter is reached by demanding the parameter $\psi$ in the Lorentz boost \eqref{locLorFinite} to scale as $\frac{\lambda}{\ell}$. Taking the $\ell\rightarrow\infty$ limit in the transformation law \eqref{locLor} of $u^a$ and $\star u^a$, we obtain (the suffix C stands for Carroll):
\begin{equation}
\label{locCar}
\delta_{\text{C}} \mu_a=\lambda  \mu^\star_a,\quad
\delta_{\text{C}}  \mu^\star_a=0,\quad
\delta_{\text{C}} v^a=0
,\quad
\delta_{\text{C}} v^a_\star=\lambda v^a,
\end{equation}
resulting in
\begin{equation}
\label{locCarths}
\delta_{\text{C}} \Xi=0,\quad 
\delta_{\text{C}} \Xi^\star=v^a\partial_a\lambda+\Xi \lambda.
\end{equation}
Using the scaling $\psi=\frac{\lambda}{\ell}$, one can compute the infinite-$\ell$ limit of equation \eqref{locLor-en-he-nc}, and deduce transformations for $\epsilon$ and $\alpha$. The result coincides exactly with \eqref{Transfo-alpha-epsilon} for 
\begin{equation}
\xi^a_{(0)}=0,\quad R=0\quad\text{and}\quad h_av^a_\star=-\lambda.
\end{equation}
This shows that \eqref{locCar}
is the ultra-relativistic version of a change of hydrodynamic frame. Finally, as explained previously, the solution space of the flat Bondi gauge is included in the flat derivative expansion, therefore it is also dual to a Carrollian fluid whose data are given in the dictionary \eqref{IndentBDE 1 v2} to \eqref{IndentBDE 2 v2}. As for anti de Sitter, the corresponding Carrollian fluid is in a particular fluid frame: the fluid velocity satisfies $v^u_\star=0$, which can be reached by acting on the derivative expansion with a finite realization of a Carrollian boost.

\section{Conclusions}\label{conclu}

We revisited in this presentation three-dimensional asymptotically locally AdS spacetimes with emphasis on the Fefferman--Graham and Bondi gauges, as well as on the fluid/gravity correspondence derivative expansion. In every instance, we described the solution space, derived the residual gauge diffeomorphisms and their action on the solution space. In particular, we showed that the solution spaces of the Fefferman--Graham and Bondi gauges are parametrized by five functions with two constrained time evolutions, while the solution space of the derivative expansion is parametrized by six functions with two constrained time evolutions. Furthermore, the residual gauge diffeomorphisms of the Fefferman--Graham and Bondi gauges are parametrized by three functions, while those of the derivative expansion are parametrized by four functions. We then showed how the gauges at hand are related to each other. The Bondi gauge can be embedded in the derivative expansion and corresponds to a particular choice of fluid frame ($u_\phi = 0$) from the point of view of the dual theory in the fluid/gravity correspondence. 
The additional parameter in the solution space of the derivative expansion is interpreted as the component $u_\phi$ of the fluid velocity, while the variations of the latter are generated by the additional parameter  $H_\phi$ in the residual gauge diffeomorphisms,   $\delta_\xi u_\phi = H_\phi$. 

Investigating the flat limit of the various items discussed within Bondi or fluid/gravity approaches was also part of our agenda.
While taking the flat limit in the Bondi gauge is straightforward, we saw that this process was more involved in the derivative expansion and required additional input regarding the $1/\ell$ behavior. We showed that the flat limit of the anti-de Sitter bulk derivative expansion gives a Ricci-flat version of it (a \emph{Carrollian-fluid/flat-gravity derivative expansion} or \emph{Ricci-flat derivative expansion} for short), along the lines of \cite{Campoleoni:2018ltl, CMPR}. The Bondi gauge in asymptotically locally flat spacetime turns out to be a sub-gauge of the Ricci-flat derivative expansion ($\mu_\phi = 0 $). Hence it inherits the corresponding holographic interpretation, as dual to a Carrollian fluid in a specific Carrollian-fluid frame.

The methods gathered in this review enjoy a promising outlook.  A natural step forward is the study of the phase space of the various gauges, both for anti de Sitter and for Minkowski.   
This would enlighten the gauge-fixing procedure in the study of asymptotic symmetries \cite{Ruzziconi:2019pzd}. The algebras of the complete sets or residual diffeomorphisms unveiled here have been worked out in \cite{CMPR},  but the computation of the charges has been performed only partially in \cite{Campoleoni:2018ltl}, for specific corners of the solution space. From this angle, the 
role played by the extra parameter in the residual diffeomorphisms of the derivative expansion, i.e. the parameter that controls the boundary local boosts (Lorentzian or Carrollian), remains to be clarified. In particular, whether this parameter generates an improper gauge transformation, eliminated when the gauge fixing is complete, is a relevant question, discussed in a more general context in  \cite{Grumiller:2016pqb , Grumiller:2017sjh}. 

Generalizing our achievements in higher dimensions is ambitious and  worth pursuing. A preliminary analysis of this line reveals that the Bondi gauge is intersecting with that of the fluid/gravity correspondence. This problem is  more challenging  than in three dimensions, where one gauge was embedded in the other, and its investigation looks appealing.

\section*{Acknowledgements}

Luca Ciambelli and Marios Petropoulos would like to thank the organizers of the \textsl{Corfu Conference on Recent Developments in Strings and Gravity} and  \textsl{Corfu Humboldt Kolleg Frontiers in Physics}, held in Corfu in September 10--19 2019, where part of the displayed material was presented. We thank Francesco Alessio, Glenn Barnich, Andrea Campoleoni, Geoffrey Comp\`ere, Laura Donnay, Adrien Fiorucci, Gaston Giribet, Victor Godet, Daniel Grumiller, Rob Leigh, Pujian Mao, Rodrigo Olea, Tasos Petkou, Kostas Siampos and C\'eline Zwikel for invaluable scientific exchange. Luca Ciambelli thanks the high-energy physics groups of the Universidad de Buenos Aires and the Universidad Andr\'es Bello, for their warm hospitality and the useful discussions on this project.  Romain Ruzziconi is a FRIA Research Fellow of the Fonds de la Recherche Scientifique F.R.S.-FNRS (Belgium). This research was supported in part by the National Science Foundation under Grant No. NSF PHY-1748958, by the Heising--Simons Foundation and by the ANR-16-CE31-0004 contract \textsl{Black-dS-String}. The work of Luca Ciambelli is supported by the ERC Advanced Grant \textsl{High-Spin-Grav}. 


\end{document}